\documentclass[12pt,a4paper]{article}
\usepackage{graphicx}
\usepackage{subeqnarray}
\begin{document}
%
%
%
%
\newenvironment{lefteqnarray}{\arraycolsep=0pt\begin{eqnarray}}
{\end{eqnarray}\protect\aftergroup\ignorespaces}
\newenvironment{lefteqnarray*}{\arraycolsep=0pt\begin{eqnarray*}}
{\end{eqnarray*}\protect\aftergroup\ignorespaces}
\newenvironment{leftsubeqnarray}{\arraycolsep=0pt\begin{subeqnarray}}
{\end{subeqnarray}\protect\aftergroup\ignorespaces}
\newcommand{\diff}{{\rm\,d}}
\newcommand{\pprime}{{\prime\prime}}
\newcommand{\szeta}{\mskip 3mu /\mskip-10mu \zeta}
\newcommand{\FC}{\mskip 0mu {\rm F}\mskip-10mu{\rm C}}
\newcommand{\appleq}{\stackrel{<}{\sim}}
\newcommand{\appgeq}{\stackrel{>}{\sim}}
\newcommand{\Int}{\mathop{\rm Int}\nolimits}
\newcommand{\Nint}{\mathop{\rm Nint}\nolimits}
\newcommand{\arcsinh}{\mathop{\rm arcsinh}\nolimits}
\newcommand{\range}{{\rm -}}
\newcommand{\displayfrac}[2]{\frac{\displaystyle #1}{\displaystyle #2}}
%
%
\title{Self-consistency and continuity questions on axisymmetric, rigidly
rotating polytropes}

\author{
 {R.~Caimmi}\footnote{
{\it Physics and Astronomy Department, Padua University,
Vicolo Osservatorio 3/2, I-35122 Padova, Italy}~~~
email: roberto.caimmi@unipd.it~~~
fax: 39-049-8278212}
\phantom{agga}}

\maketitle
\begin{quotation}
\section*{}
\begin{Large}
\begin{center}

 Abstract

\end{center}
\end{Large}
\begin{small}

\noindent\noindent
Axisymmetric, rigidly rotating polytropes are considered in the framework of
both the original Chandrasekhar (C33) approximation and a different version
(extended C33 approximation).   Special effort is devoted to two specific
points, namely (i) a contradiction between the binomial series evaluation,
$(\theta_w+\Delta\theta)^n\approx\theta_w^n+n\theta_w^{n-1}
\Delta\theta$, implying
$\vert\Delta\theta\vert\ll\vert\theta_w\vert$, and the vanishing density on
the boundary, implying $\theta_w\to0$, which affects the self-consistency of
the above mentioned approximations, and (ii) the continuity
of selected parameters as a function of the polytropic index, $n$.
Concerning (i), it is shown Emden-Chandrasekhar (EC) associated functions,
$\theta_0$, $\theta_2$, and $\psi_0$, $\psi_2$, are defined at any internal
point even if related EC
associated equations hold only for a particular subvolume, in the framework of
the extended C33 and the C33 approximation, respectively.
Concerning (ii), the continuity may safely be established in the limit,
$n\to0$, $n\to5$, for part of the parameters, while additional data are needed
for the remaining part.   Simple fitting curves, valid to a good extent for a
wide range of $n$, involve exponential functions and, in a single case, two
straight
lines joined by a parabolic segment.   The expression of physical parameters
in terms of the polytropic index can be used in building up sequences of
configurations with changing density profile for assigned 
mass and angular momentum.

\noindent

{\it keywords -
stars: equilibrium - galaxies: equilibrium - polytropes: rigid rotation.}
\end{small}
\end{quotation}

\section{Introduction} \label{intro}

A main part of astrophysical bodies, such as stars, galaxies, clusters of
galaxies, are characterized by the occurrence of three kinds of forces, due to
gravitation, rotation and pressure or stress tensor.  Density profiles are,
in principle, lying
between the extreme cases of homogeneous (constant density) and Roche (mass
point surrounded by a vanishing atmosphere) equilibrium configurations,
respectively.   A simple description implies strong restrictions, such as
homogeneity, spherical symmetry, rigid rotation.

The effect of rigid rotation on homogeneous (incompressible) ellipsoids has
been studied since a long time (e.g., MacLaurin 1742; Jacobi 1834;
Dedekind 1860; Riemann 1860; Jeans 1929, Chap.\,VIII).   For further details,
exhaustive presentation and complete references, an interested reader is
addressed to specific textbooks (e.g., Chandrasekhar 1969).

The effect of different density profiles on nonrotating (spherical) polytropes
has also been studied since a long time (e.g., Lane 1870; Ritter 1878;
Schuster 1883; Thomson 1887; Emden 1907; Chandrasekhar 1939, Chap.\,IV).   For
further
details, exhaustive presentation and complete references, an interested reader
is addressed to specific textbooks (e.g., Horedt 2004).

On the other hand, less effort has been devoted to the combined effect of
different rotation and density profiles on polytropes, due to a larger
complexity of the problem and the absence of electronic computers in the past.
When a critical rotation is attained, two different kinds of instability can
occur, namely (i) centrifugal breakup via shedding of matter on the equatorial
plane, for sufficiently steep density profiles, and (ii) 
fissional breakup after transition from axisymmetric to triaxial
configurations, for sufficiently mild density profiles.   Shedding of matter
occurs via rings or opposite streams according if related configurations are
axisymmetric or triaxial, respectively.   For further details, exhaustive
presentation and complete references, an interested reader is addressed to
specific textbooks (e.g., Jeans 1929, Chap.\,IX).

A simple and elegant first-order approximation, which implies small departure
from spherical shape, makes physical quantities depend on a rotation
parameter, $\upsilon=\Omega^2/(2\pi G\lambda)$, where $\Omega$ is the angular
velocity, $G$ the constant of gravitation, $\lambda$ the central density.
Related systems shall be quoted in the following as Emden-Chandrasekhar,
axisymmetric, rigidly rotating polytropes or, in short, EC polytropes.   For
further details, exhaustive presentation and complete references, an
interested reader is addressed to the parent paper (Chandrasekhar 1933a,
hereafter quoted as C33), and later investigations, (Chandrasekhar 1933b;
Chandrasekhar and Lebovitz 1962).

Unfortunately, the correction terms are inferred from differential equations
(hereafter quoted as EC associated equations) where the zero-th order term is
divergent on the boundary for sufficiently mild density profiles, concerning
polytropic index within the range, $0\le n<1$.   It is worth remembering that
homogeneous and extended Roche (maximal concentration: finite mass extending
up to infinity or mass point surrounded by a vanishing atmosphere within a
finite region) configurations relate to $n=0$ and $n=5$, respectively.
The above mentioned correction terms involve Legendre polynomials of degree 0
and 2, $P_0(\mu)$ and $P_2(\mu)$, respectively.

With regard to radial distortions, related to $P_0(\mu)$, the correction term,
$\chi_0$, may be incorporated into the unperturbed radial function,
$\theta_{\rm E}$, and the resulting perturbed radial function,
$\theta_0(\xi,\upsilon)=\theta_{\rm E}(\xi)+\chi_0(\xi,\upsilon)$,
is the solution of a generalized Lane-Emden
equation where rotation is also included.   Accordingly, the above mentioned
inconvenient is avoided at the price that the correction terms, $\chi_0$,
$A_2\theta_2$, depend on the rotation parameter, $\upsilon$, contrary to their
counterparts in C33 approximation, $\upsilon\psi_0$, $\upsilon C_2\psi_2$,
where the rotation parameter appears as a factor.   In the nonrotating limit,
$\upsilon\to0$, the following relations hold:
\begin{equation}
\label{eq:aflim}
\lim_{\upsilon\to0}\frac{\chi_0(\xi,\upsilon)}\upsilon=\psi_0(\xi)~~;\quad
\lim_{\upsilon\to0}\frac{A_2(\upsilon)\theta_2(\xi,\upsilon)}{C_2\upsilon}=
\psi_2(\xi)~~;
\end{equation}
where, keeping in mind both $\theta_2(\xi,\upsilon)$ and $\psi_2(\xi)$ are
undetermined by a multiplicative constant, the relation on the right-hand side
can be splitted as:
\begin{equation}
\label{eq:aflid}
\lim_{\upsilon\to0}\frac{A_2(\upsilon)}\upsilon=C_2~~;\quad
\lim_{\upsilon\to0}\theta_2(\xi,\upsilon)=\psi_2(\xi)~~;
\end{equation}
as shown below in dealing with the general theory.

The approximation under discussion shall be quoted in the following as the
extended C33 approximation.   Related systems shall be quoted as EC
polytropes, similarly to their counterparts within the framework of the C33
approximation.   For further details, exhaustive
presentation and additional references, an interested reader is addressed to
earlier attempts (Caimmi 1980, 1983, 1985, 1987, 1988;
hereafter quoted as C80, C83, C85, C87, C88, respectively).

With regard to both radial and meridional (i.e. depending on the polar angle)
distortion, the EC associated equations are formulated by use of the
power-series approximation, $\theta^n=(\theta_w+\Delta\theta)^n\approx
\theta_w^n+n\theta_w^{n-1}\Delta\theta$ which implies (C33; C83):
\begin{equation}
\label{eq:inet}
\left\vert\frac{\Delta\theta}{\theta_w}\right\vert\ll1~~;\qquad
w={\rm E},0~~;
\end{equation}
where $\theta^n$ is the distorted dimensionless
density, $\theta_w^n$ its undistorted or radially distorted counterpart,
according if $w={\rm E}$ (C33 approximation) or $w=0$ (extended C33
approximation), respectively, $\Delta\theta$ is the global
or meridional distortion term, and $\vert\Delta\theta/\theta_w\vert$ may
be considered as a distortion index.   For further details,
exhaustive presentation and complete references, an interested reader is
addressed to the parent papers (C33; C83).

The above inequality is clearly satisfied in the central region of the system,
where $\theta_w\appleq1$ and distortion due to rigid rotation is small.
Conversely, the requisite of vanishing density on the boundary of the
undistorted or radially distorted sphere, $\theta_w(\Xi_w)=0$, makes
Eq.\,(\ref{eq:inet})
violated, keeping in mind distortion due to rigid rotation is maximum on the
boundary.    Accordingly, Eq.\,(\ref{eq:inet}) is expected to hold above a
threshold, $\theta_w>\epsilon_w^\ast$, where $\epsilon_w^\ast$ depends on an
assumed tolerance, 10\% say.

The equilibrium equation of EC polytropes (hereafter quoted as EC equation),
together with the EC associated equations, have analytic solutions only in the
special cases, $n=0,1,5,$ (e.g., C33; Chandrasekhar 1939, Chap.\,IV; C80; C87;
Horedt 1990, hereafter quoted as H90; Horedt 2004, Chap.\,2).   The comparison
of related physical parameters with their counterparts, numerically computed
in a close neighbourhood of the above mentioned values of $n$, may be a useful
test for establishing the dependence on the polytropic index.

For instance,
a monotonic trend is shown by $\Xi_{\rm E}$, $\theta_{\rm E}^\prime(\Xi_
{\rm E})$, contrary to $\theta_{\rm E}^\pprime(\Xi_{\rm E})$ and EC
associated functions together with their first and second derivatives.
More specifically, the first derivatives of the EC associated functions could
be monotonic in absence of continuity at $n=0$, according to H90 results,
while the contrary holds in presence of continuity, according to C83 results,
plotted%
\footnote{
With regard to Fig.\,1 therein and related caption, $\theta_0^\prime$ has
erroneously been written instead of $\theta_0^\pprime$ and vice versa.
}
therein and listed in a later attempt (C85).   The second derivative,
$\psi_2^\pprime(\Xi_{\rm E})$, is divergent within the range, $0<n<1$.   In
conclusion, further investigation should be needed about
the continuity of physical parameters related to EC polytropes in the limit,
$n\to0$, within the framework of C33 approximation.

Concerning the opposite limit, $n\to5$, it has been noticed that, in the
nonrotating case, the dimensionless mass as a function of $n$ exhibits a
slight nonmonotonic trend with a minimum within the range, $4.80<n<4.85$,
(Seidov and Kuzakhmedov 1978).   A similar result holds for rigidly rotating
configurations (C85) which implies a slight nonmonotonic trend with a maximum,
within the above mentioned range, for the axis ratio at the onset of
equatorial breakup (C87).   Further investigation to this respect should be
needed on other physical parameters, and the continuity in the limit, $n\to5$,
should also be tested within the framework of C33 approximation.

Polytropes span over the whole range of configurations with regard not only to
density profile (from homogeneous, $n=0$, to mass point surrounded by a
vanishing atmosphere, $n=5$) and rotation (from spherical shape to fissional
or equatorial breakup), but also in connection with mechanics (classical
or relativistic) and the nature of the fluid (collisional or collisionless).
To this last respect, it has been shown that any collisional polytrope has an
exact collisionless counterpart within the range, $1/2\le n\le5$,
(Vandervoort 1980; Vandervoort and Welty 1981); which implies a description of
stellar
systems and cluster of galaxies as well, with the extension to anisotropic
stress tensors (Binney and Tremaine 1987, Chap.\,4, \S2).   For further
details, exhaustive
presentation and complete references, an interested reader is addressed to
specific textbooks (e.g., Horedt 2004).

In this view, the dependence of physical parameters on the polytropic index
could be useful for a number of applications even if $n\ge1/2$ for
collisionless systems.   For instance, liquid cores within planets or more
exotic objects, such as neutron stars and quark stars, and deep oceans
within satellites, could be described as polytropes with low $n$.   Quasi
static contraction via energy dissipation where total mass and angular
momentum are left unchanged, while the polytropic index is increasing, could
be described provided the dependence of selected physical parameters on $n$ is
known.

The isopycnic (i.e. constant density) surfaces can be approximated as similar
and similarly placed ellipsoids (exact for homogeneous configurations) for
several investigations, such as the description of
gravitational radiation from collapsing and rotating massive star cores,
(Saens and Shapiro 1978, 1981), rigidly rotating and binary
polytropes, (Lai et al. 1993, 1994a,b), gravitational collapse of nonbaryonic
dark matter and related pancake formation (Bisnovatyi-K\"ogan 2004, 2005),
kinetic energy of ellipsoidal matter distributions (Rodrigues 2014).
In short, inhomogeneous configurations can be described, to an acceptable
extent, in terms of properties related to homogeneous configurations.

A description of tenuous gas-dust atmospheres of some stars
and tenuous haloes surrounding compact elliptical galaxies, in terms of
extended Roche configurations, is mentioned in a recent investigation
(Kondratyev and Trubitsina 2013).

On the other hand, for reasons outlined above, the homogeneous $(n=0)$ and
extended Roche $(n=5)$ limit for polytropes have never been attained (to the
knowledge of the author) using numerical simulations where, at most, $0.1\le n
\le4.9$.   To this respect, a first step must necessarily be performed
analytically.

The current attempt is restricted to the investigation of two specific points,
namely (i) the extent to which, for different density profiles and rotation
rates, the C33 approximation (both in its original and extended form) is
self-consistent in the sense that
Eq.\,(\ref{eq:inet}) is satisfied for an assumed tolerance equal to 10\%, and
(ii) the dependence on the polytropic index, shown by five physical
parameters, which are
selected in order to avoid divergence at the limiting configurations, $n=0$
(homogeneous) and $n=5$ (extended Roche system).   More specifically, one
among the above mentioned parameters is related to the nonrotating
configuration, three to the rotating configuration, one to the onset of
equatorial breakup.

The paper is organized as follows.   The general theory of rigidly rotating
polytropes is briefly outlined in Section \ref{geth}.   The dependence of the
distortion index, $\vert\Delta\theta/\theta_w\vert$, on the polytropic
index, $n$, $0\le n\le5$, the dimensionless radial coordinate, $\xi$,
$0\le\xi\le\Xi_{\rm E}$, and the rotation parameter, $\upsilon$,
$0\le\upsilon\le\upsilon_{\rm R}$, is determined in Section \ref{secm}.   Five
selected physical parameters are plotted as a function of the polytropic
index, $n$, in Section \ref{parn}, where simple fitting functions are
determined and relative errors are shown.   The discussion is performed in
Section \ref{disc}.   The conclusion is drawn in Section \ref{conc}.
Further analysis on slowly rotating isopycnic surfaces and detailed
exposition of the fitting procedure are left to the Appendix.

\section{General theory}
\label{geth}

The theory of rigidly rotating polytropes has been exhaustively developed in
earlier attempts (e.g., Jeans 1929, Chap.\,IX; C33; C80; C83; Horedt 2004,
Chap.\,3);
and shall not be repeated here, leaving aside extensions and improvements.
An interested reader is addressed to the above quoted parent papers.   Only
what is relevant for the current investigation shall be reviewed in the
following.

\subsection{EC and EC associated equations}
\label{EC}

In dimensionless coordinates, the EC equation reads:
\begin{leftsubeqnarray}
\slabel{eq:ECa}
&& \frac1{\xi^2}\frac\partial{\partial\xi}\left(\xi^2\frac{\partial
\theta}{\partial\xi}\right)+
\frac1{\xi^2}\frac\partial{\partial\mu}\left[(1-\mu^2)\frac{\partial
\theta}{\partial\mu}\right]-\upsilon=-\theta^n;\quad \\
\slabel{eq:ECb}
&& \theta(0,\mu)=1~~;\quad\left(\frac{\partial\theta}{\partial\xi}\right)_
{0,\mu}=0~~;\quad\left(\frac{\partial\theta}{\partial\mu}\right)_{0,\mu}=0~~;
\quad\\
\slabel{eq:ECc}
&& \lim_{\upsilon\to0}\theta(\xi,\mu)=\theta_{\rm E}(\xi_{\rm E})~~;
\label{seq:EC}
\end{leftsubeqnarray}
where $\xi$ is a dimensionless radial distance, $\delta=\arccos\mu$ the angle
with respect to the rotation axis (polar angle), $n$ the polytropic index
$(0\le n\le5)$, $\theta^n$ a dimensionless density, $\upsilon$ a dimensionless
rotation parameter, and the index, E, denotes nonrotating configurations
$(\upsilon=0)$.

The usual physical quantities relate to their dimensionless counterparts as:
\begin{lefteqnarray}
\label{eq:csial}
&& r=\alpha\xi~~;\qquad\alpha=\left[\frac{(n+1)p_{\rm c}}{4\pi G\lambda^2}
\right]^{1/2}=\left[\frac{(n+1)K\lambda^{1/n}}{4\pi G\lambda}\right]^{1/2}~~;
\quad\\
\label{eq:pirho}
&& p=K\rho^{1+1/n}~~;\qquad\rho=\lambda\theta^n~~; \\
\label{eq:upsi}
&& \upsilon=\frac{\Omega^2}{2\pi G\lambda}~~;
\end{lefteqnarray}
where $r$ is the radial distance, $\alpha$ a scaling radius, $p$ the pressure,
$p_{\rm c}$ the central pressure, $\rho$ the density, $\lambda$ the central
density, $K$ related to the central temperature, $G$ the gravitation constant
and $\Omega$ the angular velocity.   For further details, an interested reader
is addressed  to earlier attempts (e.g., C33; C80).

The general solution to the EC equation, Eq.\,(\ref{eq:ECa}), hereafter quoted
as EC function, can be expanded in series of Legendre polynomials as:
\begin{lefteqnarray}
\label{eq:thL}
&& \theta(\xi,\mu)=\sum_{\ell=0}^{+\infty}A_{2\ell}\theta_{2\ell}(\xi)P_
{2\ell}(\mu)~~;
\end{lefteqnarray}
where odd terms are ruled out by symmetry with respect to the equatorial plane
and $A_{2\ell}$ are coefficients which, for $2\ell>0$, depend on the rotation
parameter.

The boundary conditions related to the EC associated functions,
$\theta_{2\ell}$, can be inferred from
Eq.\,(\ref{eq:ECb}) via (\ref{eq:pirho}) and (\ref{eq:thL}).   The result is:
\begin{lefteqnarray}
\label{eq:bc2l}
&& \theta_{2\ell}(0)=\delta_{0,2\ell};\quad \theta_{2\ell}^\prime(0)=0;
\quad\lim_{\upsilon\to0}\theta_0(\xi)=\theta_{\rm E}(\xi_{\rm E});\quad \\
\label{eq:A2ls}
&& A_0=1~~;\qquad\lim_{\upsilon\to0}A_{2\ell}(\upsilon)=0~~;\qquad2\ell>0~~;
\end{lefteqnarray}
where $\delta_{ij}$ is the Kronecker symbol, not to be confused with the
polar angle, $\delta=\arccos\mu$.

The Legendre polynomials, $P_{\ell}(\mu)$, can be expressed as:
\begin{lefteqnarray}
\label{eq:Lp}
&& P_\ell(\mu)=\frac1{2^\ell}\frac1{\ell!}\frac{\diff^\ell}{\diff\mu^\ell}
[(\mu^2-1)^\ell]~~;
\qquad\vert P_\ell(\mu)\vert\le1~~;\qquad\ell=0,1,2,...~~;\qquad
\end{lefteqnarray}
which satisfy the Legendre equations:
\begin{lefteqnarray}
\label{eq:Le}
&& \frac{\diff}{\diff\mu}\left[(1-\mu^2)\frac{\diff P_\ell}{\diff\mu}\right]=
-\ell(\ell+1)P_\ell(\mu)~~;
\end{lefteqnarray}
for non negative integer $\ell$.   For further details, an interested reader
is addressed  to classical textbooks on the theory of the potential (e.g.,
MacMillan 1930, Chap.\,VII, \S\S185-192).

If the distortion due to rigid rotation may be considered as a small
perturbation with respect to the spherical shape, then the first term of the
series expansion on the right-hand side of Eq.\,(\ref{eq:thL}) is dominant
and the power on the right-hand side of the EC equation, Eq.\,(\ref{eq:ECa}),
can safely be approximated as:
\begin{lefteqnarray}
\label{eq:thn2l}
&& [\theta(\xi,\mu)]^n=[\theta_0(\xi)]^n
+n[\theta_0(\xi)]^{n-1}\sum_{\ell=1}^
{+\infty}A_{2\ell}\theta_{2\ell}(\xi)P_{2\ell}(\mu)~~; \\
\label{eq:ep0}
&& \vert\theta_0(\xi)\vert>\epsilon_0^\ast~~;
\end{lefteqnarray}
which is a series expansion in Legendre polynomials, provided a fixed
threshold, $\epsilon_0^\ast$, is not exceeded.

The substitution of Eqs.\,(\ref{eq:thL}) and (\ref{eq:thn2l}) into the EC
equation, Eq.\,(\ref{eq:ECa}), taking separately the terms of same degree in
Legendre polynomials, yields:
\begin{lefteqnarray}
\label{eq:EC2l}
&& \frac1{\xi^2}\frac\diff{\diff\xi}\left(\xi^2\frac{\diff
\theta_{2\ell}}{\diff\xi}\right)-\frac{2\ell(2\ell+1)}{\xi^2}\theta_{2\ell}
=\frac{\delta_{0,2\ell}}{A_{2\ell}}(-\theta_0^n+\upsilon)-(1-\delta_{0,2\ell})
n\theta_0^{n-1}\theta_{2\ell};\qquad
\end{lefteqnarray}
that is the EC associated equations of degree, $2\ell$.   If
$\theta_0<0$, the real part of the principal value of the complex power,
$(\theta_0)^x$, has to be considered.   For further details and exhaustive
presentation, an interested reader is addressed to earlier attempts
(Linnel 1981; C83; Geroyannis 1988; Geroyannis and Karageorgopoulos 2014).

\subsection{Isopycnic surfaces}
\label{issu}

When a spherical polytrope attains rigid rotation, mass elements outside the
rotation axis are displaced further away, yielding an oblate configuration
where the polar axis coincides with the rotation axis.   More specifically,
oblateness weakens the gravitational force along the polar axis and
strenghtens the gravitational force along the equatorial plane, but the
gravitational + centrifugal force is also weakened, which implies loss of
spherical shape.

Let $\theta_{\rm E}(\xi_{\rm E})=\kappa$ and $\theta(\xi,\mu)=\kappa$ be a
generic isopycnic surface related to the nonrotating
and rigidly rotating configuration, respectively.   In terms of the
dimensionless radial coordinate, $\xi$, the isopycnic surface can be expressed
as $\xi=\xi_{\rm E}$ and $\xi=\xi(\mu)$, respectively.   Let the polar and the
equatorial coordinate be denoted as $\xi_p=\xi(1)$ and $\xi_e=\xi(0)$,
respectively, where oblateness implies $\xi_p\le\xi_{\rm E}\le\xi_e$.

The generic oblate isopycnic surface, via Eqs.\,(\ref{eq:thL}),
(\ref{eq:bc2l}), (\ref{eq:A2ls}), reads:
\begin{lefteqnarray}
\label{eq:iso}
&& \theta(\xi,\mu)=\theta_0(\xi)+R_1(\xi,\mu)=\kappa~~; \\
\label{eq:R1}
&& R_1(\xi,\mu)=\sum_{\ell=1}^{+\infty}A_{2\ell}\theta_{2\ell}(\xi)P_{2\ell}
(\mu)~~;
\end{lefteqnarray}
where $\xi_p\le\xi\le\xi_e$ owing to oblateness.   In addition,  the following
inequality:
\begin{lefteqnarray}
\label{eq:inen}
&& \vert R_1(\xi,\mu)\vert\ll\epsilon_0^\ast<\vert\theta_0(\xi)\vert~~;
\end{lefteqnarray}
implies the validity of Eq.\,(\ref{eq:thn2l}).
The middle side of Eq.\,(\ref{eq:iso})
describes an expansion of the nonrotating polytrope as a whole, via
$\theta_0$, and superimposed on this an oblateness, via $R_1$.   More
specifically, $\theta_0$ relates to an expanded sphere, where the radial
contribution of rigid rotation adds to the undistorted configuration, and
$R_1$ quantifies the meridional distortion.

\begin{figure*}[t]  
\begin{center}      
\includegraphics[scale=0.8]{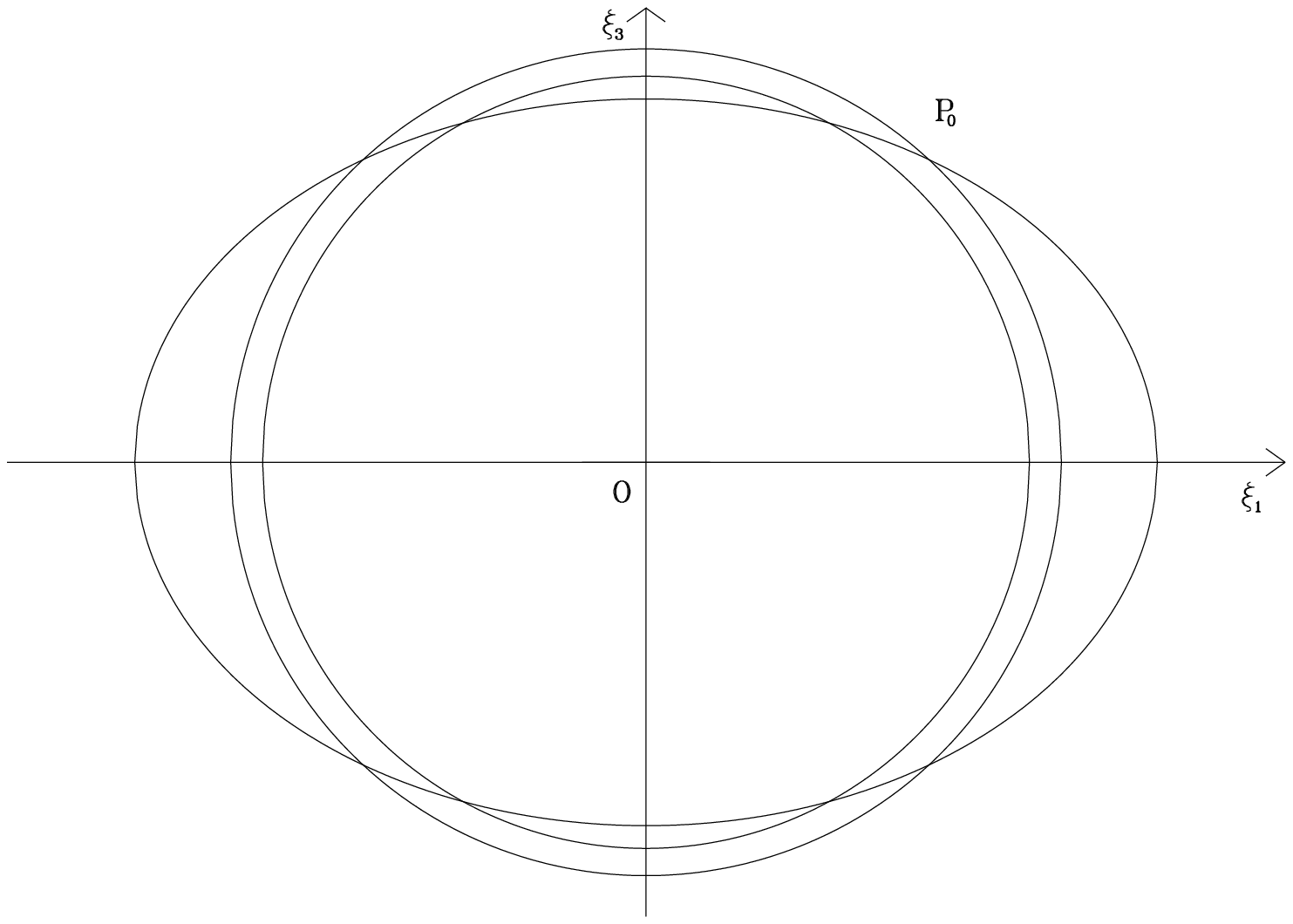}                      
\caption[ddbb]{Generic isopycnic surfaces related to an assigned polytrope in
the nonrotating limit (inner spherical), $\theta_{\rm E}(\xi_{\rm E})=\kappa$,
and in rigid rotation (oblate), $\theta(\xi,\mu)=\kappa$.
The (fictitious) isopycnic surface of the expanded sphere (outer spherical),
$\theta_0(\xi_0)=\kappa$, is also shown.   Outer spherical and oblate
isopycnic surfaces intersect at the locus, $(\xi_0,\mp\mu_0)$, where
$\mu_0=\cos\delta_0$, $\delta$ polar angle, e.g.,
${\sf P}_0\equiv[\vert(\xi_0)_1\vert,\vert(\xi_0)_3\vert]$,
$(\xi_0)_1=\xi_0\sqrt{1-\mu_0^2}$, $(\xi_0)_3=\mp\xi_0\mu_0$.   For
further details refer to the text.}
\label{f:iso}     
\end{center}       
\end{figure*}

Let $\xi_0$ define the (fictitious) isopycnic surface of the expanded sphere,
as:
\begin{lefteqnarray}
\label{eq:is0}
&& \theta_0(\xi_0)=\theta_{\rm E}(\xi_{\rm E})=\kappa~~;
\end{lefteqnarray}
where radial expansion implies $\xi_{\rm E}\le\xi_0$.   The substitution of
Eq.\,(\ref{eq:is0}) into (\ref{eq:iso}) yields: 
\begin{lefteqnarray}
\label{eq:R10}
&& R_1(\xi_0,\mp\mu_0)=0~~;
\end{lefteqnarray}
and the locus, $(\xi_0,\mp\mu_0)$, defines the intersection
between the oblate isopycnic surface, $\theta(\xi,\mu)=\kappa$, and the
(fictitious) isopycnic surface, $\theta_0(\xi_0)=\kappa$, as depicted in
Fig.\,\ref{f:iso}.
In the nonrotating limit, $\upsilon\to0$, $\xi_0\to\xi_{\rm E}$, and the term
containing $P_2(\mu)$ is expected to be dominant with respect to the others in
Eq.\,(\ref{eq:R1}).   Accordingly, $\mu_0$ relates to $P_2(\mu)=0$, hence
$\mu_0\to1/\sqrt{3}$.   For further details, an interested reader is addressed
to Appendix \ref{a:P2}.

\subsection{Gravitational potential}
\label{grapo}

The gravitational potential within EC polytropes reads (e.g.,
C80):
\begin{equation}
\label{eq:potg}
{\cal V}_{\rm G}(\xi,\mu)=4\pi G\lambda\alpha^2\left\{\theta(\xi,\mu)-\frac16
\upsilon\xi^2[1-P_2(\mu)]\right\}+{\cal V}_{\rm p}~~;
\end{equation}
where ${\cal V}_{\rm p}$ is an additive constant
which, at the moment, remains undetermined.

The substitution of Eq.\,(\ref{eq:thL}) into (\ref{eq:potg}) yields:
\begin{lefteqnarray}
\label{eq:po2l}
&& {\cal V}_{\rm G}(\xi,\mu)=4\pi G\lambda\alpha^2\sum_{\ell=0}^{+\infty}
\left[A_{2\ell}\theta_{2\ell}(\xi)-\frac{\delta_{2\ell,0}-\delta_{2\ell,2}}6
\upsilon\xi^2\right]
P_{2\ell}(\mu)+{\cal V}_{\rm p}~~;
\end{lefteqnarray}
where $A_0=1$ according to Eq.\,(\ref{eq:A2ls}).

The gravitational potential of a body of revolution,
at sufficiently large distance outside the boundary, can be expressed as
(e.g., MacMillan 1930, Chap.\,VII, \S193):
\begin{equation}
\label{eq:pex}
{\cal V}_{\rm G}(\xi,\mu)=4\pi G\lambda\alpha^2\sum_{\ell=0}^{+\infty}\frac
{c_{2\ell}}{\xi^{2\ell+1}}P_{2\ell}(\mu)~~;\qquad\xi\gg\Xi~~;
\end{equation}
where $c_{2\ell}$ are dimensionless coefficients
and the odd terms are ruled out by symmetry with respect to the equatorial
plane.

For points near the boundary, Eq.\,(\ref{eq:pex}) is exact only for
spherical-symmetric matter distributions and Roche systems, but remains
acceptable provided
oblateness maintains sufficiently small or concentration maintains
sufficiently high.   The worst case relates to homogeneous configurations
$(n=0)$ where, on the other hand, the gravitational potential may be expressed
analytically (e.g., MacMillan 1930, Chap.\,II, \S39).   The best case relates
to extended Roche systems $(n=5)$, where either the boundary is
infinitely distant from the centre of mass or the whole mass is concentrated
at a single point surrounded by a massless atmosphere filling a finite volume,
both implying an exact formulation.

The continuity of the gravitational potential and the gravitational force on
a selected point of the boundary, $\Xi=\Xi(\mu)$, implies
Eqs.\,(\ref{eq:po2l}), (\ref{eq:pex}), and related first
derivatives, match at $(\Xi,\mu)$ for the terms of the same degree in
Legendre polynomials.   The result is:
\begin{lefteqnarray}
\label{eq:rpot}
&& A_{2\ell}\theta_{2\ell}(\Xi)-\frac{\delta_{2\ell,0}-\delta_{2\ell,2}}6
\upsilon\Xi^2+\delta_{2\ell,0}c_p=\frac{c_{2\ell}}{\Xi^{2\ell+1}}~~; \\
\label{eq:rfor}
&& A_{2\ell}\theta_{2\ell}^\prime(\Xi)-\frac{\delta_{2\ell,0}-
\delta_{2\ell,2}}3\upsilon\Xi=-\frac{(2\ell+1)c_{2\ell}}{\Xi^{2\ell+2}}~~; \\
\label{eq:cp}
&& c_p=\frac{{\cal V}_p}{4\pi G\lambda\alpha^2}~~;
\end{lefteqnarray}
where $A_0=1$, Eq.\,(\ref{eq:A2ls}), and the constants, $A_{2\ell}$,
$c_{2\ell}$, $c_p$, are the solutions of the system, Eqs.\,(\ref{eq:rpot}) and
(\ref{eq:rfor}), for $2\ell=0, 2, 4, ...$.

After performing a lot of algebra, related explicit expressions read:
\begin{lefteqnarray}
\label{eq:cpe}
&& c_p=-\theta_0(\Xi)-\Xi\theta_0^\prime(\Xi)+\frac12\upsilon\Xi^2~~; \\
\label{eq:c0}
&& c_0=-\Xi^2\theta_0^\prime(\Xi)+\frac13\upsilon\Xi^3~~;
\end{lefteqnarray}
for $2\ell=0$ where, in general, $\theta_0(\Xi)=\kappa_b$, and:
\begin{lefteqnarray}
\label{eq:A2}
&& A_2=-\frac56\frac{\upsilon\Xi^2}{3\theta_2(\Xi)+\Xi\theta_2^\prime(\Xi)}~~; \\
\label{eq:c2}
&& c_2=-\frac\upsilon6\frac{\Xi^5[2\theta_2(\Xi)-\Xi\theta_2^\prime(\Xi)]}{3
\theta_2(\Xi)+\Xi\theta_2^\prime(\Xi)}
=\frac15A_2\Xi^3
[2\theta_2(\Xi)-\Xi\theta_2^\prime(\Xi)]~~;
\end{lefteqnarray}
for $2\ell=2$, and:
\begin{lefteqnarray}
\label{eq:A2l}
&& A_{2\ell}[(2\ell+1)\theta_{2\ell}(\Xi)+\Xi\theta_{2\ell}^\prime(\Xi)]=0
~~;
\end{lefteqnarray}
which implies the following:
\begin{lefteqnarray}
\label{eq:A2l0}
&& A_{2\ell}=0~~;\qquad c_{2\ell}=0~~;
\end{lefteqnarray}
for $2\ell>2$.   More specifically, a null value of the sum within square
brackets in Eq.\,(\ref{eq:A2l}) would be in contradiction with the EC
associated equation, Eq.\,(\ref{eq:EC2l}).

Accordingly, Eqs.\,(\ref{eq:iso}), (\ref{eq:R1}), (\ref{eq:inen}), reduce to:
\begin{lefteqnarray}
\label{eq:is2}
&& \theta(\xi,\mu)=\theta_0(\xi)+A_2\theta_2(\xi)P_2(\mu)=\kappa~~; \\
\label{eq:R2}
&& R_1(\xi,\mu)=A_2\theta_2(\xi)P_2(\mu)~~; \\
\label{eq:ine2}
&& \vert A_2\theta_2(\xi)P_2(\mu)\vert\ll\epsilon_0^\ast<\vert\theta_0(\xi)
\vert~~;
\end{lefteqnarray}
where, in particular, $\kappa=\kappa_{\rm b}$ on the boundary.   In addition, 
Eq.\,(\ref{eq:R10}) reduces to:
\begin{lefteqnarray}
\label{eq:R20}
&& A_2\theta_2(\xi_0)P_2(\mp\mu_0)=0~~;
\end{lefteqnarray}
which implies $\mu_0=1/\sqrt{3}$, $\delta_0=\arctan\sqrt2$, and the locus of
intersections between isopycnic surfaces, $\theta(\xi,\mu)=\kappa$ and
$\theta_0(\xi_0)=\kappa$, is the surface of a cone with axis coinciding with
the polar axis, vertex coinciding with the centre of mass and generatrixes,
$\delta_0=\arctan\sqrt2$.

If, in particular, $\Xi=\Xi_p$ in Eqs.\,(\ref{eq:cpe})-(\ref{eq:A2l}), the
equation of the boundary via Eq.\,(\ref{eq:is2}) specifies to:
\begin{lefteqnarray}
\label{eq:Csip}
&& \theta(\Xi_p,1)=\theta_0(\Xi_p)+A_2\theta_2(\Xi_p)P_2(1)=\kappa_{\rm b}~~;
\end{lefteqnarray}
where $P_2(1)=1$.   Then the combination of Eqs.\,(\ref{eq:A2}) and
(\ref{eq:Csip}) yields:
\begin{lefteqnarray}
\label{eq:A2t}
&& \frac56\frac{\upsilon\Xi_p^2}{3\theta_2(\Xi_p)+\Xi_p\theta_2^\prime(\Xi_p)}
=\frac{\theta_0(\Xi_p)-\kappa_{\rm b}}{\theta_2(\Xi_p)}~~;
\end{lefteqnarray}
which is a transcendental equation in $\Xi_p$.   Keeping in mind
$\theta_2(\xi)\sim\xi^2$ as $\xi\to0$ (C33; C83) and $\theta_0(\Xi_0)=
\kappa_{\rm b}$,
$\Xi_0\ge\Xi_p$, the left-hand side of Eq.\,(\ref{eq:A2t}) maintains both
finite and positive, while the right-hand side is monotonically decreasing
from positive infinite to zero via Eqs.\,(\ref{eq:bc2l}) and (\ref{eq:is0}),
within the range, $0\le\Xi_p\le\Xi_0$.   Accordingly, Eq.\,(\ref{eq:Csip})
admits a unique solution within the above mentioned range.

The dimensionless equatorial semiaxis, $\Xi_e$, can be determined using the
equation of the boundary via Eq.\,(\ref{eq:is2}), which translates into:
\begin{lefteqnarray}
\label{eq:Csie}
&& \theta(\Xi_e,0)=\theta_0(\Xi_e)+A_2\theta_2(\Xi_e)P_2(0)=\kappa_{\rm b}~~;
\end{lefteqnarray}
where $P_2(0)=-1/2$, which is a transcendental equation in $\Xi_e$.   The
knowledge of the dimensionless semiaxes, $\Xi_p$, $\Xi_e$, implies the
knowledge of the axis ratio, as:
\begin{lefteqnarray}
\label{eq:eps}
&& \epsilon=\frac{\alpha\Xi_p}{\alpha\Xi_e}=\frac{\Xi_p}{\Xi_e}~~;
\end{lefteqnarray}
according to Eq.\,(\ref{eq:csial}).

In the following, the whole procedure for determining $A_2$ and related
quantities via Eq.\,(\ref{eq:A2}), particularized to $\Xi=\Xi_p$, shall be
quoted as the Chandrasekhar procedure (C33) extended to the pole of the
system or, in short, the extended C33 procedure.

If the isopycnic surfaces are expressed by Eq.\,(\ref{eq:is2}), a better
method for determining the constants, $c_p$ and $A_2$, acts as follows: (1)
calculate the gravitational potential at the centre of mass using the
equilibrium equation via Eq.\,(\ref{eq:potg}) and the mass distribution via
Eq.\,(\ref{eq:pirho}), and express $c_p$ by comparison of related results; (2)
calculate the gravitational force at the pole using the
equilibrium equation via Eq.\,(\ref{eq:potg}) and the mass distribution via
Eq.\,(\ref{eq:pirho}), and express $A_2$ by comparison of related results.
For further details, an interested reader is addressed  to to the parent
papers (C80; C83).

In the following, the whole procedure for determining $A_2$ and related
quantities, as outlined above, shall be quoted as the C80 procedure.

\subsection{C33 approximation}
\label{C33}

A simpler approximation was used in the original parent paper (C33).   More
specifically, Eq.\,(\ref{eq:thL}) is expressed therein as:
\begin{lefteqnarray}
\label{eq:thC}
&& \theta(\xi,\mu)=\theta_{\rm E}(\xi)+\upsilon R_0(\xi,\mu)~~; \\
\label{eq:R0}
&& R_0(\xi,\mu)=\sum_{\ell=0}^{+\infty}C_{2\ell}\psi_{2\ell}(\xi)
P_{2\ell}(\mu)~~; \\
\label{eq:psbc}
&& C_0=1~~;\qquad\psi_{2\ell}(0)=0~~;\qquad\psi_{2\ell}^\prime(0)=0~~;
\end{lefteqnarray}
where $C_{2\ell}$ are constants and $\xi_p\le\xi\le\xi_e$ on a selected
isopycnic surface.

If the distortion due to rigid rotation may be considered as a small
perturbation with respect to the spherical shape, then the first term of the
series expansion on the right-hand side of Eq.\,(\ref{eq:thC}) is dominant
and the power on the right-hand side of the EC equation, Eq.\,(\ref{eq:ECa}),
can safely be approximated as:
\begin{lefteqnarray}
\label{eq:thC2l}
&& [\theta(\xi,\mu)]^n=[\theta_{\rm E}(\xi)]^n+n[\theta_{\rm E}(\xi)]^{n-1}
\upsilon
\sum_{\ell=0}^{+\infty}C_{2\ell}\psi_{2\ell}(\xi)P_{2\ell}(\mu)
~~; \\
\label{eq:epE}
&& \vert\theta_{\rm E}(\xi)\vert>\epsilon_{\rm E}^\ast~~;
\end{lefteqnarray}
which is a series expansion in Legendre polynomials, provided a fixed
threshold, $\epsilon_{\rm E}^\ast$, is not exceeded.

The substitution of Eqs.\,(\ref{eq:thC}) and (\ref{eq:thC2l}) into the EC
equation, Eq.\,(\ref{eq:ECa}), taking separately the terms of same degree in
Legendre polynomials, yields:
\begin{lefteqnarray}
\label{eq:CC2l}
&& \frac1{\xi^2}\frac\diff{\diff\xi}\left(\xi^2\frac{\diff
\psi_{2\ell}}{\diff\xi}\right)-\frac{2\ell(2\ell+1)}{\xi^2}\psi_{2\ell}
=\delta_{2\ell,0}-n\theta_{\rm E}^{n-1}\psi_{2\ell}~~;\quad
\end{lefteqnarray}
where $\psi_0$ and $\psi_{2\ell}$, $2\ell>0$, relate to the radial expansion
and the meridional distortion, respectively, due to rigid rotation.   The
comparison between Eqs.\,(\ref{eq:EC2l}) and (\ref{eq:CC2l}) discloses that:
\begin{lefteqnarray}
\label{eq:psig0}
&& \psi_0(\xi)=\lim_{\upsilon\to0}\frac{\theta_0(\xi,\upsilon)-\theta_{\rm E}
(\xi)}\upsilon~~; \\
\label{eq:psin0}
&& \lim_{\upsilon\to0}[\theta_0(\xi,\upsilon)]^n=\lim_{\upsilon\to0}
[\theta_{\rm E}(\xi)+\upsilon\psi_0(\xi)]^n
\nonumber \\
&& \phantom{\lim_{\upsilon\to0}[\theta_0(\xi,\upsilon)]^n}
=\lim_{\upsilon\to0}\{[\theta_{\rm E}(\xi)]^n+n[\theta_{\rm E}(\xi)]^{n-1}
\upsilon\psi_0(\xi)\}~~;~\quad \\
\label{eq:psig2}
&& C_{2\ell}\psi_{2\ell}(\xi)=\lim_{\upsilon\to0}\frac{A_{2\ell}(\upsilon)
\theta_{2\ell}(\xi,\upsilon)}{\upsilon}~~;\quad2\ell>0~~;
\end{lefteqnarray}
which enlightens the difference between C33 and extended C33 approximation.

Keeping in mind the solutions of EC associated equations,
Eqs.\,(\ref{eq:EC2l}) and (\ref{eq:CC2l}), remain undetermined by a
multiplicative constant for $2\ell>0$, Eq.\,(\ref{eq:psig2}) may be splitted
as:
\begin{lefteqnarray}
\label{eq:psig3}
&& C_{2\ell}=\lim_{\upsilon\to0}\frac{A_{2\ell}(\upsilon)}\upsilon~~;\quad
\psi_{2\ell}(\xi)=\lim_{\upsilon\to0}\theta_{2\ell}(\xi,\upsilon)~~;\quad
2\ell>0~~;\quad
\end{lefteqnarray}
with no loss of generality.

The gravitational potential and the gravitational force inside and outside
the system can be matched on the boundary of the nonrotating sphere, $\Xi=\Xi_
{\rm E}$, and Eqs.\,(\ref{eq:cpe})-(\ref{eq:A2l0}) still hold provided $\Xi_p$,
$\theta_0$, $A_{2\ell}\theta_{2\ell}$, are replaced by $\Xi_{\rm E}$,
$\theta_{\rm E}+\upsilon\psi_0$, $\upsilon C_{2\ell}\psi_{2\ell}$,
respectively.   The result is:
\begin{lefteqnarray}
\label{eq:dpE}
&& c_p=c_{p{\rm E}}+\upsilon d_p~;\quad c_{p{\rm E}}=-\theta_{\rm E}
(\Xi_{\rm E})-\Xi_{\rm E}\theta_{\rm E}^\prime(\Xi_{\rm E})~; \nonumber \\
&& d_p=-\psi_0(\Xi_{\rm E})-\Xi_{\rm E}\psi_0^\prime(\Xi_{\rm E})+\frac12
\Xi_{\rm E}^2~~; \\
\label{eq:cE}
&& c_0=c_{\rm E}+\upsilon d_0~~;\quad c_{\rm E}=-\Xi_{\rm E}^2\theta_{\rm E}^
\prime(\Xi_{\rm E})~~; \nonumber \\
&& d_0=-\Xi_{\rm E}^2\psi_0^\prime(\Xi_{\rm E})+\frac13\Xi_{\rm E}^3~~; \\
\label{eq:C2}
&& A_2=\upsilon C_2~~;\quad C_2=-\frac56\frac{\Xi_{\rm E}^2}
{3\psi_2(\Xi_{\rm E})+\Xi_{\rm E}\psi_2^\prime(\Xi_{\rm E})}~~; \\
\label{eq:d2}
&& c_2=\upsilon d_2~;\quad d_2=
\frac15C_2\Xi_{\rm E}^3[2\psi_2(\Xi_{\rm E})-\Xi_{\rm E}\psi_2^\prime
(\Xi_{\rm E})]~;\qquad \\
\label{eq:d2l}
&& C_{2\ell}=0~~;\quad d_{2\ell}=0~~;\quad2\ell>2~~;
\end{lefteqnarray}
for further details, an interested reader is addressed to the parent paper
(C33).

In the following, the whole procedure for determining $A_2=\upsilon C_2$ and
related quantities by use of the Chandrasekhar approximation (C33) shall be
quoted as the C33 procedure.

\section{A self-consistent method}
\label{secm}

The validity of the EC associated equations, Eq.\,(\ref{eq:EC2l}), via
Eq.\,(\ref{eq:thn2l}), is implied by the inequality, expressed by
Eq.\,(\ref{eq:ine2}).  For a generic isopycnic surface, defined by
Eq.\,(\ref{eq:is2}), the above mentioned inequality can be formulated as:
\begin{lefteqnarray}
\label{eq:dita}
&& \zeta(\xi,\mu)\ll1~~; \\
\label{eq:zita}
&& \zeta(\xi,\mu)=\left\vert\frac{A_2\theta_2(\xi)P_2(\mu)}{\theta_0(\xi)}\right
\vert=\left\vert\frac{\theta_0(\xi_0)}{\theta_0(\xi)}-1\right\vert~~;
\end{lefteqnarray}
where $\theta_0(\xi_0)=\kappa$ and $\xi_0=\xi(\mp1/\sqrt3)$ is the
intersection between a selected oblate isopycnic surface and its (fictitious)
counterpart related to the expanded sphere.   The function,
$\zeta(\xi,\mu)$, may be conceived as a distortion indicator.   In particular,
$\zeta(\xi_0,\mu_0)=0$ and $\zeta(\Xi,\mu)=1$ provided $\Xi\ne\Xi_0$ and
$\kappa_{\rm b}=0$.   For
fixed $\xi$, the distortion indicator has a maximum on the polar axis,
$\zeta(\xi,\mu)\le\zeta(\xi_p,1)=\zeta_p(\xi_p)$, due to $P_2(\mu)\le1$.
According to Eq.\,(\ref{eq:zita}), the distortion indicator, $\zeta_p$, is
maximized as $\kappa_{\rm b}=\theta_0(\Xi_0)\to0$ or $\theta(\Xi,\mu)\to0$
which, on the other hand, is mostly considered in literature (e.g.,
Horedt 2004, Chap.\,2, \S2.2).   For this reason, $\kappa_{\rm b}=0$ shall be
assumed in the following.   To
get deeper insight, special cases for which analytical solutions exist shall
first be considered.

\subsection{The special case $n=0$}
\label{n0}

The solutions of the EC associated equations, Eqs.\,(\ref{eq:EC2l}),
(\ref{eq:CC2l}), for $n=0$, keeping in mind Eqs.\,(\ref{eq:psig0}),
(\ref{eq:psig3}), read:
\begin{lefteqnarray}                                              
\label{eq:th00}
&& \theta_0(\xi)=1-\frac{1-\upsilon}6\xi^2~~; \\
\label{eq:Csi00}
&& \Xi_0=\frac{\Xi_{\rm E}}{(1-\upsilon)^{1/2}}=\left(\frac6{1-\upsilon}
\right)^{1/2}~~; \\
\label{eq:psi00}
&& \psi_0(\xi)=\lim_{\upsilon\to0}\frac{\theta_0(\xi)-\theta_{\rm E}(\xi)}
\upsilon=\frac16\xi^2~~; \\
\label{eq:th20}
&& \theta_2(\xi)=\psi_2(\xi)=\xi^2~~;
\end{lefteqnarray}
where the boundary conditions are expressed by Eq.\,(\ref{eq:bc2l}).

According to the extended C33 procedure, Eq.\,(\ref{eq:A2}) reduces to:
\begin{lefteqnarray}
\label{eq:A20}
&& A_2=-\frac16\upsilon~~;
\end{lefteqnarray}
and the substitution of Eqs.\,(\ref{eq:th00})-(\ref{eq:A20}) into
(\ref{eq:Csip})-(\ref{eq:eps}) yields after some algebra:
\begin{lefteqnarray}
\label{eq:assi0}
&& \Xi_p=\Xi_{\rm E}~;\quad\Xi_e=\Xi_{\rm E}\left(1-\frac32\upsilon\right)^
{-1/2}~;\quad\Xi_{\rm E}=\sqrt6~; \nonumber \\
&& \epsilon=\left(1-\frac32\upsilon\right)^{1/2}~;\qquad
\end{lefteqnarray}
accordingly, the rotation parameter, $\upsilon$, the dimensionless radius of
the expanded
sphere, $\Xi_0$, and the coefficient, $A_2$, may be expressed in terms of the
axis ratio, $\epsilon$, as:
\begin{lefteqnarray}
\label{eq:upep0}
&& \upsilon=(1-\epsilon^2)\gamma(1)~~; \\
\label{eq:C0ep0}
&& \Xi_0=\Xi_{\rm E}\left(\frac3{1+2\epsilon^2}\right)^{1/2}~~; \\
\label{eq:A2ep0}
&& A_2=-\frac{1-\epsilon^2}6\gamma(1)~~;
\end{lefteqnarray}
where $\gamma(1)=2/3$, according to the next Eq.\,(\ref{eq:gamma}).
The above results cannot be considered as acceptable approximations, leaving
aside the dependence of $A_2$ on $\epsilon$, Eq.\,(\ref{eq:A2ep0}), which
shows an increasing discrepancy for decreasing $\epsilon$ up to 30\% for
$\epsilon=0$ (C80).

An exact formulation can be derived following the C80 procedure.   The result
is:
\begin{lefteqnarray}
\label{eq:A2s}
&& A_2=-\frac{1-\epsilon^2}6\gamma(\epsilon)~~;\qquad\epsilon=\frac{\Xi_p}
{\Xi_e}~~; \\
\label{eq:gamma}
&& \gamma(\epsilon)=\frac2{1-\epsilon^2}\left[1-\epsilon\frac{\arcsin\sqrt{1-
\epsilon^2}}{\sqrt{1-\epsilon^2}}\right]~~; \nonumber \\
&& \gamma(1)=\frac23~;\qquad\gamma(0)=2~;\qquad \\
\label{eq:dgam}
&& \frac{\diff\gamma}{\diff\epsilon}=\frac1{1-\epsilon^2}\left[(1+2\epsilon^2)
\frac\gamma\epsilon-\frac2\epsilon\right]~~; \\
\label{eq:vs}
&& \upsilon(\epsilon)=1-\frac{1+2\epsilon^2}2\gamma(\epsilon)~~;\qquad\upsilon
(1)=\upsilon(0)=0~~;
\end{lefteqnarray}
accordingly, $\upsilon(\epsilon)$ is nonmonotonic and $A_2(1)=0$, $A_2(0)=
-1/3$.

The coordinates of the extremum point, which must necessarily be positive,
satisfy the transcendental equation, $\diff\upsilon/\diff\epsilon=0$, or
(e.g., Caimmi 2006):
\begin{lefteqnarray}
\label{eq:exup}
&& \gamma(\epsilon)=2\frac{1+2\epsilon^2}{1+8\epsilon^2}~~;
\end{lefteqnarray}
which has two solutions  on the boundary of the domain, according to
Eq.\,(\ref{eq:gamma}), and a third solution inside the domain, related to the
extremum point (e.g., Chandrasekhar 1969, Chap.\,5, \S32).
The flat configuration, $\epsilon=0$, $\upsilon=0$, is also centrifugally
supported, $\upsilon_{\rm R}=0$, but the system is unstable towards bar modes
for $\epsilon<0.58272$.   For further details and additional references, an
interested reader is addressed to the parent paper (C80).

The substitution of Eqs.\,(\ref{eq:th00})-(\ref{eq:th20}) and (\ref{eq:A2ep0})
into (\ref{eq:zita}) yields after some algebra:
\begin{lefteqnarray}
\label{eq:zit0}
&& \zeta(\xi,\mu)=\left\vert\frac{\xi^2-\xi_0^2}{\Xi_0^2-\xi^2}\right\vert~~;
\end{lefteqnarray}
where it can be seen $\zeta(\xi,\mu)\ll1$ implies $\xi_0\ll\Xi_0$ inside
the related isopycnic surface of the expanded sphere, $\xi<\xi_0$, and $\xi\ll
[(\Xi_0^2+\xi_0^2)/2]^{1/2}$ outside the related isopycnic surface of the
expanded sphere but inside the expanded sphere, $\xi_0<\xi<\Xi_0$, while no
acceptable solution to the above mentioned inequality exists outside the
expanded sphere, $\xi>\Xi_0$.

For practical purposes, it is better dealing with the upper limit,
$\zeta_p(\xi_p)=\zeta(\xi_p,1)\ge\zeta(\xi,\mu)$.   To this respect, the
substitution of
Eqs.\,(\ref{eq:th00}), (\ref{eq:th20}) and (\ref{eq:A20}) or (\ref{eq:A2s})
into (\ref{eq:zita}) yields, after some algebra, an explicit expression of the
distortion indicator, $\zeta_p$, for the extended C33 or the C80 procedure,
respectively.

\subsection{The special case $n=1$}
\label{n1}

The solutions of the EC associated equations, Eqs.\,(\ref{eq:EC2l}),
(\ref{eq:CC2l}), for $n=1$, keeping in mind Eqs.\,(\ref{eq:psig0}),
(\ref{eq:psig3}), read%
\footnote{
It is worth noticing the associated EC function, $\theta_2(\xi)$, is lacking
of the factor, 15, in the parent paper (C80) which, on the other hand, is
absorbed by the constant, $A_2~(B_2$ therein), leaving the resutls unchanged.
Conversely, the factor, 15, appears in a subsequent attempt (C83).
}:
\begin{lefteqnarray}
\label{eq:th01}
&& \theta_0(\xi)=(1-\upsilon)\frac{\sin\xi}\xi+\upsilon~~; \\
\label{eq:psi01}
&& \psi_0(\xi)=\lim_{\upsilon\to0}\frac{\theta_0(\xi)-\theta_{\rm E}(\xi)}
\upsilon=1-\frac{\sin\xi}{\xi}~~; \\
\label{eq:th21}
&& \theta_2(\xi)=\psi_2(\xi)=15\left[\left(\frac3{\xi^2}-1\right)\frac
{\sin\xi}\xi-\frac3\xi\frac{\cos\xi}\xi\right]~~;
\end{lefteqnarray}
where the boundary conditions are expressed by Eq.\,(\ref{eq:bc2l}).   The
first derivative of the last function, after some algebra, reads:
\begin{lefteqnarray}
\label{eq:dt21}
&& \theta_2^\prime(\xi)=\psi_2^\prime(\xi) \nonumber \\
&& \phantom{\theta_2^\prime(\xi)}
=15\left[\left(\frac9{\xi^3}+\frac4
\xi\right)\frac{\sin\xi}\xi+\left(\frac9{\xi^2}-1\right)\frac{\cos\xi}\xi
\right]~~;
\end{lefteqnarray}
and the substitution of Eqs.\,(\ref{eq:th21}), (\ref{eq:dt21}), into
(\ref{eq:A2}), according to the extended C33 procedure, after some algebra
yields:
\begin{lefteqnarray}
\label{eq:A21}
&& A_2=-\frac{\upsilon\Xi^2}{18}\left[\left(\frac{18}{\Xi^2}+1
\right)\frac{\sin\Xi}{\Xi}-\cos\Xi\right]^{-1}~~;
\end{lefteqnarray}
and the dimensionless semiaxes, $\Xi_p$, $\Xi_e$, are the solution of the
transcendental equations, Eqs.\,(\ref{eq:A2t}), (\ref{eq:Csie}), respectively,
via (\ref{eq:th01}), taking $\kappa_{\rm b}=0$.

Numerical computations must be performed following the C80 procedure, as the
gravitational potential within the system cannot be expressed analytically.
It can be seen the rotation parameter, $\upsilon$, and the coefficient,
$-A_2$, monotonically increase up to about 0.107 and 0.962, respectively, as
the axis ratio, $\epsilon$, decreases up to about 0.531.   Additional rotation
would imply equatorial breakup.   For further details, an interested reader is
addressed to the parent paper (C80).

\subsection{The special case $n=5$}
\label{n5}

The solutions of the EC associated equations, Eqs.\,(\ref{eq:EC2l}),
(\ref{eq:CC2l}), for $n=5$, keeping in mind Eqs.\,(\ref{eq:psig0}),
(\ref{eq:psig3}), read:
\begin{lefteqnarray}
\label{eq:th05}
&& \theta_0(\xi)=\cos\nu+\frac12\upsilon\tan^2\nu~~;\qquad\nu=\arctan\frac\xi
{\sqrt3}~~; \\
\label{eq:psi05}
&& \psi_0(\xi)=\lim_{\upsilon\to0}\frac{\theta_0(\xi)-\theta_{\rm E}(\xi)}
\upsilon=\frac12\tan^2\nu~~; \\
\label{eq:th25}
&& \theta_2(\xi)=\psi_2(\xi)=\frac{15}{128}\left\{\frac{3[\nu-\sin\nu\cos^3\nu
+\sin^3\nu\cos\nu]}{\sin^3\nu\cos^2\nu}\right. \nonumber \\
&& \phantom{\theta_2(\xi)=\psi_2(\xi)=}-\left.
\frac{8[\sin^3\nu\cos^5\nu-\sin^5\nu\cos^3\nu]}{\sin^3\nu\cos^2\nu}\right\};
\quad
\end{lefteqnarray}
where the boundary conditions are expressed by Eq.\,(\ref{eq:bc2l}).

With regard to Eqs.\,(\ref{eq:th05})-(\ref{eq:psi05}),
it can be seen $\theta_{\rm E}(\xi)=\cos\nu$ fits to the correct nonrotating
limit (e.g., C80) while $\psi_0(\xi)=(1/2)\tan^2\nu$ is an approximate
solution of Eq.\,(\ref{eq:CC2l}) which attains the same asymptotic expression
as the exact
solution, $\psi_0(\xi)\to\xi^2/6$, for both $\xi\to0$ and $\xi\to+\infty$
(H90).   On the other hand, $\upsilon\to0$ in the case under
discussion and $\upsilon\psi_0(\xi)$ is infinitesimal while
$\theta_{\rm E}(\xi)$ maintains finite provided $\xi<+\infty$.   Accordingly,
Eq.\,(\ref{eq:th05}) may be considered as infinitely close to the exact
solution of the EC associated equation of degree, $2\ell=0$,
Eq.\,(\ref{eq:EC2l}).

In the limit, $\xi\to+\infty$, the following relations hold:
\begin{lefteqnarray}
\label{eq:limf0}
&& \lim_{\xi\to+\infty}\left[\xi\theta_0(\xi)-\frac16\upsilon\xi^3\right]=
\sqrt3~~;\qquad \\
\label{eq:limf01}
&& \lim_{\xi\to+\infty}\left[\xi^2\theta_0^\prime(\xi)-\frac13\upsilon\xi^3
\right]=-\sqrt3~~;\qquad \\
\label{eq:limf1}
&& \lim_{\xi\to+\infty}\left[\xi^3\theta_0^\pprime(\xi)-\frac13\upsilon\xi^3
\right]=2\sqrt3~~;\qquad \\
\label{eq:limf12}
&& \lim_{\xi\to+\infty}\left[\xi^{-2}\theta_2(\xi)\right]=\frac{15\pi}{256}
~~;\qquad \\
\label{eq:limf2}
&& \lim_{\xi\to+\infty}\left[\xi^{-1}\theta_2^\prime(\xi)\right]=\frac{15\pi}
{128}~~;\qquad \\
\label{eq:limf2ab}
&& \lim_{\xi\to+\infty}\theta_2^\pprime(\xi)=\frac{15\pi}{128}~~;\qquad \\
\label{eq:limf2a}
&& \lim_{\xi\to+\infty}\left[2\theta_2(\xi)-\xi\theta_2^\prime(\xi)\right]=
\frac{225\pi}{256}~~;\qquad
\end{lefteqnarray}
where the dimensionless boundary, $\Xi(\mu)$, extends to infinity and
$\kappa_{\rm b}=0$.

In the nonrotating limit, $\upsilon\to0$, $\theta_0(\xi)\to\theta_{\rm E}
(\xi)$, Eqs.\,(\ref{eq:limf0})-(\ref{eq:limf1}) reduce to: 
\begin{lefteqnarray}
\label{eq:limE0}
&& \lim_{\xi\to+\infty}\left[\xi\theta_{\rm E}(\xi)\right]=\sqrt3~~;\qquad \\
\label{eq:limE01}
&& \lim_{\xi\to+\infty}\left[\xi^2\theta_{\rm E}^\prime(\xi)\right]=-\sqrt3~~;
\qquad \\
\label{eq:limE1}
&& \lim_{\xi\to+\infty}\left[\xi^3\theta_{\rm E}^\pprime(\xi)\right]=2\sqrt3
~~;\qquad
\end{lefteqnarray}
where the dimensionless boundary, $\Xi_{\rm E}(\mu)$, attains infinity.

According to the extended C33 procedure, Eqs.\,(\ref{eq:cpe})-(\ref{eq:c2})
reduce to:
\begin{lefteqnarray}
\label{eq:A25}
&& c_p=0~~;\quad c_0=\sqrt3~~;\quad A_2=-\frac{128}{45\pi}\upsilon~~;\quad
c_2=-\frac12\upsilon\Xi^3~~;\quad
\end{lefteqnarray}
and the substitution of Eqs.\,(\ref{eq:th05})-(\ref{eq:A25}) into
(\ref{eq:Csip})-(\ref{eq:eps}) after some algebra yields:
\begin{lefteqnarray}
\label{eq:assi5}
&& \Xi_p=\Xi_{\rm E}~~;\qquad\Xi_e=\frac{\Xi_{\rm E}}\epsilon~~; \\
\label{eq:upE5}
&& \upsilon\Xi_{\rm E}^3=4\sqrt3\epsilon^2(1-\epsilon)~~; \\
\label{eq:upe5}
&& \upsilon\Xi_e^3=4\sqrt3\frac{1-\epsilon}\epsilon~~;
\end{lefteqnarray}
where, in particular, centrifugal equilibrium is attained at the equator for
$\upsilon\Xi_e^3=2\sqrt3$, which implies $\epsilon=2/3$, $\upsilon_{\rm R}\Xi_
{\rm E}^3=16\sqrt3/27$.   Additional rotation
makes the onset of equatorial breakup.   For further details, an interested
reader is addressed to the parent paper (C87).

The substitution of Eqs.\,(\ref{eq:th05})-(\ref{eq:th25}) and (\ref{eq:A25})
into (\ref{eq:zita}), keeping in mind $\epsilon\ge2/3$, yields after some
algebra:
\begin{lefteqnarray}
\label{eq:zit5}
&& \zeta(\xi,\mu)=\left\vert\frac{\sqrt3/\xi_0+\upsilon\xi_0^2/6}
{\sqrt3/\xi+\upsilon\xi^2/6}\right\vert~~;
\end{lefteqnarray}
for infinite dimensionless distances, $\xi\to+\infty$; if otherwise,
$\zeta(\xi,\mu)\to0$.

For practical purposes, it is better dealing with the upper limit,
$\zeta_p(\xi_p)=\zeta(\xi_p,1)\ge\zeta(\xi,\mu)$.   To this respect, the
substitution of
Eqs.\,(\ref{eq:limf0}), (\ref{eq:limf12}), (\ref{eq:A25})-(\ref{eq:upe5}),
into (\ref{eq:zita}) yields after some algebra:
\begin{lefteqnarray}
\label{eq:zip5}
&& \zeta_p(\xi_p)=\zeta(\xi_p,1)=\left\{1+\left[\frac23\epsilon^2(1-\epsilon)
\frac{\xi_p}{\Xi_{\rm E}}\right]^{-1}\right\}^{-1}~~;
\end{lefteqnarray}
provided $\xi_p\to+\infty$.  The special case of centrifugal equilibrium,
$\epsilon=2/3$, at the pole, $\xi_p=\Xi_p=\Xi_{\rm E}$, reads
$\zeta(\Xi_{\rm E},1)=8/89$.

The above results can be obtained following the C80 procedure which, in the
case under discussion, is equivalent to the extended C33 procedure, in that
both are exact.   For further details, an interested reader is addressed to
the parent paper (C87).

\subsection{The general case}
\label{n}

Equilibrium configurations related to both C33 and extended C33 procedure are
increasingly near to their exact counterparts as the rotation parameter is
increasingly smaller, $\upsilon\to0$.   On the other hand, equilibrium
configurations related to the C80 procedure are conceptually exact but suffer
from approximations related to numerical integrations.   In particular,
$\upsilon$ is infinitesimal for $n=5$, where the isopycnic surfaces are
distorted only at infinite dimensionless distances from the centre of mass,
$\xi\to+\infty$, and the extended C33 procedure yields exact results.

Conversely, $\upsilon$ rises up to a maximum and later decreases
for $n=0$, where the isopycnic surfaces (intended as coinciding
with gravitational + centrifugal equipotential surfaces for
homogeneous configurations) can be distorted, leaving aside
instabilities, up to a flat disk, and the extended C33
procedure yields very rough results unless $\upsilon\to0$, $\epsilon\to1$.

The special case,
$n=1$, lies between the above mentioned extreme situations, $n=5$ and $n=0$.
Accordingly, results from both the C33 and the extended C33 procedure are
expected to provide a
better description of equilibrium configurations for increasing polytropic
index, $n$.

Aiming to a rapid insight to the dependence of the distortion indicator,
$\zeta_p$, on
the polytropic index, $n$, and the rotation parameter, $\upsilon$, the EC
associated functions, $\theta_0$, $\theta_2$, shall be expressed according to
the extended C33 procedure, but approximated according to the C33 procedure.
The result is:
\begin{lefteqnarray}
\label{eq:psi0}
&& \theta_0(\xi,\upsilon)=\theta_{\rm E}(\xi)+\upsilon\psi_0(\xi)~~; \\
\label{eq:psi2}
&& A_2\theta_2(\xi,\upsilon)=\upsilon C_2\psi_2(\xi)~~; \\
\label{eq:Apsi2}
&& C_2=\frac{A_2}\upsilon~~;
\end{lefteqnarray}
and the substitution of Eqs.\,(\ref{eq:is2}), (\ref{eq:psi0}),
(\ref{eq:psi2}), into (\ref{eq:zita}) yields after some algebra:
\begin{lefteqnarray}
\label{eq:zitE}
&& \zeta(\xi,\mu)\leq\zeta(\xi_p,1)=\zeta_p(\xi_p)=\left\vert\frac{\upsilon
C_2\psi_2(\xi)}{\theta_{\rm E}(\xi)+\upsilon\psi_0(\xi)}\right\vert
\ll1~~;
\end{lefteqnarray}
which may readily be calculated in that the functions, $\theta_{\rm E}$,
$\psi_0$, $\psi_2$, are tabulated.

On the boundary of the nonrotating sphere, $\xi=\Xi_{\rm E}$, $\theta_{\rm E}
(\Xi_{\rm E})=0$, high-precision values of $\Xi_{\rm E}$, $\theta_{\rm E}^
\prime(\Xi_{\rm E})$, $\psi_0(\Xi_{\rm E})$, $\psi_0^\prime(\Xi_{\rm E})$,
$\psi_2(\Xi_{\rm E})$, $\psi_2^\prime(\Xi_{\rm E})$, are available (e.g.,
C85; H90).   In addition, $\zeta_p(\Xi_{\rm E})=-C_2
\psi_2(\Xi_{\rm E})/\psi_0(\Xi_{\rm E})$, regardless of the rotation
parameter, $\upsilon$, with the exception of the special case, $n=5$, where
$\theta_{\rm E}(\xi)$, $\upsilon\psi_0(\xi)$, $\upsilon\psi_2(\xi)$, are
infinitesimal of the same order for $\xi\to\Xi_{\rm E}$, as can be inferred
from Eqs.\,(\ref{eq:limf0})-(\ref{eq:limf2ab}).   High-precision values of
$\theta_{\rm E}(\xi)$,
$0\le\xi\le\Xi_{\rm E}$, are also available (e.g., Horedt 1986).   On the
other hand, to the knowledge of the author, values of $\psi_0(\xi)$,
$\psi_0^\prime(\xi)$, $\psi_2(\xi)$, $\psi_2^\prime(\xi)$, $0<\xi<\Xi_
{\rm E}$, can be found only in the parent paper (C33).    The above mentioned
references make the source of the data used for analysing the dependence of
the distortion indicator, $\zeta_p(\xi_p)$, on the polytropic index, $n$,
the rotation parameter, $\upsilon$, and the fractional dimensionless radial
coordinate, $\xi/\Xi_{\rm E}$.

\begin{figure*}[t]  
\begin{center}      
\includegraphics[scale=0.8]{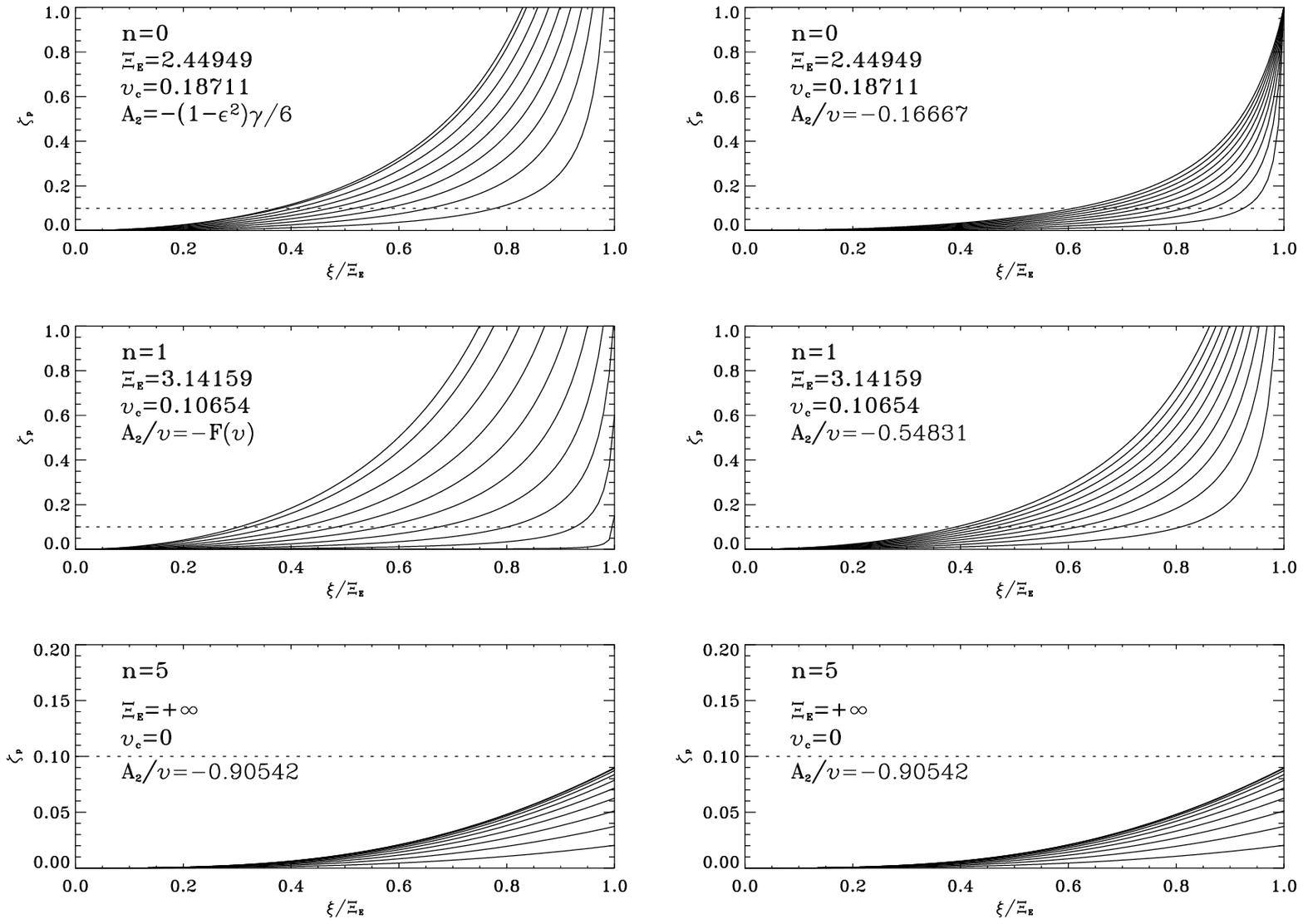}                      
\caption[ddbb]{The distortion indicator, $\zeta_p$, against the fractional
dimensionless radial coordinate, $\xi/\Xi_{\rm E}$, for polytropic index
values, $n=0,1,5$.   Different curves relate to different rotation parameter
values, $\upsilon/\upsilon_c=0.1,0.2,...,1.0$, from bottom to top, where
$\upsilon_c$ is a critical value denoting the onset of instability against bar
modes $(n\appleq0.808)$ or equatorial breakup $(n\appgeq0.808)$.   Left and
right panels relate to the C80 and C33 procedure, respectively, which are
coincident in the special case, $n=5$, where the vertical scale is enlarged.
For further details refer to the text.}
\label{f:zita6}     
\end{center}       
\end{figure*}

\begin{figure*}[t]  
\begin{center}      
\includegraphics[scale=0.8]{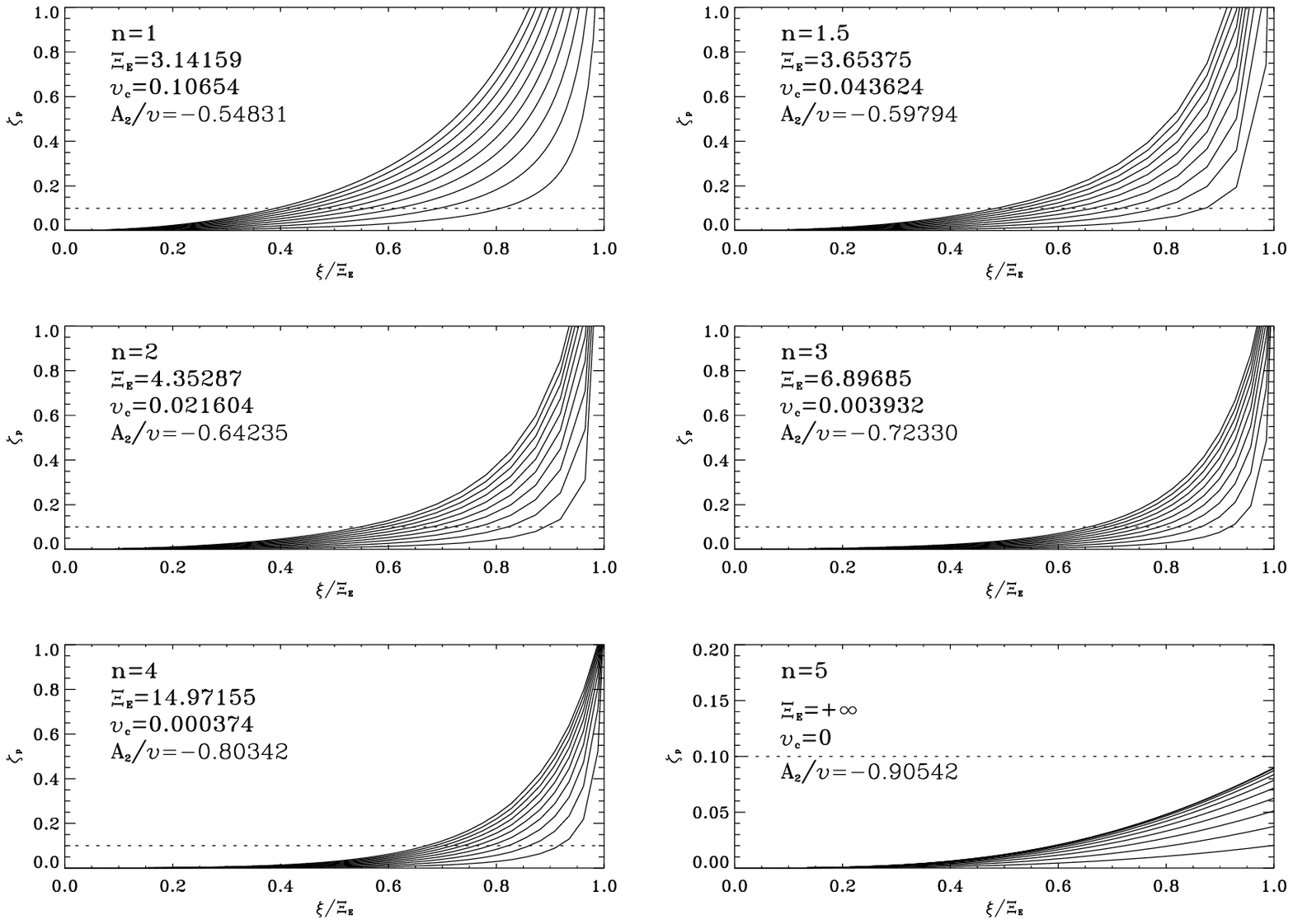}                      
\caption[ddbb]{As in Fig.\,\ref{f:zita6} but for $n=1.5,2,3,4$, according to
the C33 procedure.   Additional cases already plotted in 
Fig.\,\ref{f:zita6}, $n=1,5$, are added for better comparison.}
\label{f:zitb6}     
\end{center}       
\end{figure*}

The results are shown in Figs.\,\ref{f:zita6} and \ref{f:zitb6}, where
different curves (from bottom to top) relate to rotation parameter values,
$\upsilon/\upsilon_c=0.1,0.2,...,1.0$, and $\upsilon_c$ is a
threshold which denotes the onset of instability against bar modes
$(n\appleq0.808)$ or equatorial breakup $(n\appgeq0.808)$.   To gain insight,
cases $n=1,5,$ are repeated and the special value, $\zeta_p=0.1$, is marked by
a dashed horizontal straight line.   
More specifically, left and right panels of Fig.\,\ref{f:zita6} correspond to
the C80 and C33 procedure, respectively, while only the latter is
considered in Fig.\,\ref{f:zitb6}.   All curves diverge at $\xi=\Xi_0$,
$\theta_0(\Xi_0)=0$, according to Eq.\,(\ref{eq:zita}), provided $\mu\ne\mu_0$
and $n<5$.    The parameter, $C_2=A_2/\upsilon$, is independent of the
rotation parameter, $\upsilon$, according to Eq.\,(\ref{eq:C2}), in the C33
procedure, while the contrary holds, according to Eqs.\,(\ref{eq:A2s}),
(\ref{eq:vs}),
for $n=0$, and to numerical results for $0<n<5$, in the C80 procedure.   In
the special case, $n=5$, both the C33 and the C80 procedure coincide with
the exact theory.

It is worth noticing critical rotation parameter values are taken from
different investigations.   While there is general consensus on the onset of
barlike instabilities (in absence of tidal potential) at $\upsilon_c=0.1871$
for $n=0$
(e.g., Jeans 1929, Chap.\,VIII, \S192), the contrary holds for the onset of
equatorial breakup, as shown in Table \ref{t:uc}.
\begin{table*}
\caption[par]{Values of rotation parameter at the onset of equatorial breakup,
$\upsilon_{\rm R}$, for
different values of polytropic index, $n$, taken from earlier
investigations: James 1964 (J64), Hurley and Roberts 1964 (HR64), Monaghan and
Roxburgh 1965 (MR65),
Martin 1970 (M70), Naylor and Anand 1970 (NA70), C80, Horedt 1983 (H83), C85.
The decimal notation after $\upsilon_{\rm R}$ values, E$-i$, $i$ natural
number, is represented as -$i$ to save space.}
\label{t:uc}
\begin{center}
\begin{tabular}{lllllllll}
\hline
\multicolumn{1}{c}{$n$} &
\multicolumn{8}{c}{$\upsilon_{\rm R}$} \\
\multicolumn{1}{c}{} &
\multicolumn{8}{c}
{\_\_\_\_\_\_\_\_\_\_\_\_\_\_\_\_\_\_\_\_\_\_\_\_\_\_\_\_\_\_\_\_\_\_\_\_\_\_\_\_\_\_\_\_\_\_\_\_\_\_\_\_\_\_\_\_\_\_\_\_\_\_\_\_\_\_\_\_\_\_\_\_\_\_\_\_\_\_\_\_\_\_\_\_\_\_\_\_\_\_} \\
\multicolumn{1}{c}{} &
\multicolumn{1}{c}{J64} &
\multicolumn{1}{c}{HR64} &
\multicolumn{1}{c}{MR65} &
\multicolumn{1}{c}{M70} &
\multicolumn{1}{c}{NA70} &
\multicolumn{1}{c}{C80} &
\multicolumn{1}{c}{H83} &
\multicolumn{1}{c}{C85} \\
\hline\noalign{\smallskip}
0.0   &                          &        &         &         &         & 0           &        & 6.6667-1 \\
0.808 & 1.060296\hspace{0.0mm}-1 &        &         &         &         &             & 1.22-1 & 1.3323-1 \\
1.0   & 8.3720  \hspace{2.7mm}-2 &        & 7.59-2  &         &         & 1.0654$ $-1 & 9.46-2 & 1.2040-1 \\
1.5   & 4.3624  \hspace{2.7mm}-2 & 4.45-2 & 4.10-2  & 4.16-2  & 3.75-2  &             & 4.80-2 & 5.1942-2 \\
2.0   & 2.1604  \hspace{2.7mm}-2 &        & 1.99-2  & 2.14-2  & 1.94-2  &             & 2.34-2 & 2.4964-2 \\
2.5   & 9.9300  \hspace{2.7mm}-3 & 1.01-2 & 9.31-3  & 9.90-3  &         &             & 1.07-2 & 1.1513-2 \\
3.0   & 3.932   \hspace{4.8mm}-3 & 4.13-3 & 3.95-3  & 4.08-3  & 3.93-3  &             & 4.36-3 & 4.6946-3 \\
3.5   &                          & 1.40-3 & 1.25-3  &         &         &             & 1.48-3 & 1.5842-3 \\
4.0   &                          & 3.33-4 & 3.27-4  & 3.29-4  & 3.22-4  &             & 3.50-4 & 3.7434-4 \\
4.75  &                          &        &         &         &         &             &        & 3.9697-6 \\
4.9   &                          &        &         & 2.03-7  & 2.03-7  &             & 2.88-7 &          \\
5.0   &                          &        &         &         &         & 0           &        & 0        \\
\noalign{\smallskip}
\hline
\end{tabular}
\end{center}
\end{table*}
For $n=1$, values from C80 were necessarily to be used in that values of
$A_2(\upsilon)$ were also used in dealing with the C80 procedure.   Related
middle panels of Fig.\,\ref{f:zita6} can be renormalized to the J64
(James 1964) value, $(\upsilon_c)_{\rm J64}=0.083720$, keeping in mind 
$0.8(\upsilon_c)_{\rm C80}=0.8\times0.10654=0.085232$, which implies the last
upper two curves must be neglected for closer comparison with cases,
$n=1.5,2,3$.   For $n=4$, $\upsilon_c$ has been taken from a different source
(C85).
A full list of values together with related references can be found
in specific textbooks (e.g., Horedt 2004, Chap.\,3, \S3.8.8) with the addition
of a few exceptions (e.g., C80) and recent attempts (e.g., Geroyannis and
Karageorgopoulos 2014).

An inspection of Figs.\,\ref{f:zita6} and \ref{f:zitb6} discloses the
following.
\begin{description}
\item[$\bullet$\hspace{6mm}]
For small values of the rotation parameter,
$\upsilon/\upsilon_c\approx0.1$-$0.2$, the distortion indicator satisfies
$\zeta_p\appleq0.1$ within the range of polytropic index, $0\le n<5$,
provided $\xi/\Xi_{\rm E}\appleq0.8$.
\item[$\bullet$\hspace{6mm}]
For large values of the rotation parameter,
$\upsilon/\upsilon_c\approx0.8$-$1.0$, the distortion indicator satisfies
$\zeta_p\appleq0.1$ within the range, $0\le n\appleq1$, provided
$\xi/\Xi_{\rm E}\appleq0.4$; within the range, $1\appleq n\appleq2$, provided
$\xi/\Xi_{\rm E}\appleq0.5$; within the range, $2\le n<5$, provided
$\xi/\Xi_{\rm E}\appleq0.6$.
\item[$\bullet$\hspace{6mm}]
For all values of the rotation parameter,
$0\le\upsilon/\upsilon_c\le1$, the distortion indicator satisfies
$\zeta_p\appleq0.1$ in the special case, $n=5$, within the range,
$\xi/\Xi_{\rm E}\le1$.
\end{description}
Let $\zeta_p\appleq0.1$ be assumed as a criterion for the validity, to an
acceptable extent, of Eq.\,(\ref{eq:thn2l}).    According to the above
results, it holds up to $\xi/\Xi_{\rm E}\appleq0.8$ in the limit of low
rotation parameter, $\upsilon/\upsilon_c\le0.1$, and roughly up to
$\xi/\Xi_{\rm E}\approx(n+8)/20$ in the limit of large rotation parameter,
$0.9\le\upsilon/\upsilon_c\le1$, provided $n<5$, while it holds up to
$\xi/\Xi_{\rm E}=1$, regardless of the rotation parameter, provided $n=5$.
Outside the above mentioned ranges in $\xi/\Xi_{\rm E}$, Eq.\,(\ref{eq:thn2l})
no longer holds and the EC equation, Eq.\,(\ref{eq:ECa}), should be integrated
using a different kind of approximation, unless $n=0,1,5$.

\section{Dependence on $n$ of selected parameters}
\label{parn}

The investigation of the dependence on the polytropic index, $n$, has been
restricted to the following parameters: $c_{\rm E}/\Xi_{\rm E}$,
$d_0/\Xi_{\rm E}$, $d_2/\Xi_{\rm E}^3$, $C_2$,
$\upsilon_{\rm R}\Xi_{\rm E}^3$,
which are related to well defined physical features, namely the continuity of
the gravitational potential on the boundary and, concerning the last one, to
the onset of equatorial breakup.   As a first step, let the above mentioned
functions be explicitly expressed in the special cases, $n=0, 1, 5$.

With regard to the special case, $n=0$, the substitution of
Eqs.\,(\ref{eq:th00})-(\ref{eq:th20}) into (\ref{eq:cE})-(\ref{eq:d2}) after
some algebra yields:
\begin{lefteqnarray}
\label{eq:par0}
&& \frac{c_{\rm E}}{\Xi_{\rm E}}=2~~;\quad\frac{d_0}{\Xi_{\rm E}}=0~~;\quad
\frac{d_2}{\Xi_{\rm E}^3}=0~~;\quad~~C_2=-\frac16~~;\quad
n=0~~;
\end{lefteqnarray}
where the onset of equatorial breakup, $\upsilon_{\rm R}=0$, $\epsilon_{\rm R}
=0$, as already mentioned, relates to an unstable configuration.

With regard to the special case, $n=1$, the substitution of
Eqs.\,(\ref{eq:th01})-(\ref{eq:th21}) into (\ref{eq:cE})-(\ref{eq:d2}) after
some algebra yields:
\begin{lefteqnarray}
\label{eq:par1}
&& \frac{c_{\rm E}}{\Xi_{\rm E}}=1~;~~\frac{d_0}{\Xi_{\rm E}}=\frac{\pi^2}3-1
~;~~\frac{d_2}{\Xi_{\rm E}^3}=-\frac{15-\pi^2}3~;~~
C_2=-\frac{\pi^2}{18}~;~~n=1~;\qquad
\end{lefteqnarray}
where the value of rotation parameter at the onset of equatorial breakup,
$\upsilon_{\rm R}$, cannot be analytically expressed and depends on the
method used, as shown in Table \ref{t:uc}.

With regard to the special case, $n=5$, the substitution of
Eqs.\,(\ref{eq:th05})-(\ref{eq:th25}) into (\ref{eq:cE})-(\ref{eq:d2}) after
some algebra yields:
\begin{lefteqnarray}
\label{eq:par5}
&& \frac{c_{\rm E}}{\Xi_{\rm E}}=0~;~~\frac{d_0}{\Xi_{\rm E}}=0~;~~
\frac{d_2}{\Xi_{\rm E}^3}=-\frac12~;~~C_2=-\frac{128}{45\pi}~;
~~n=5~;\qquad
\end{lefteqnarray}
where $\psi_0(\xi)\sim\xi^2/6$ as $\xi\to+\infty$ (H90) and the exact
expression for finite $\xi$ (H90) does not matter as $\upsilon\to0$ in the
case under consideration, $\upsilon\psi_o(\xi)\to0$ for finite $\xi$, which
makes the effect of distorsion negligible therein (C87).   At the onset of
equatorial breakup, $\epsilon_{\rm R}=2/3$, $\upsilon_{\rm R}\Xi_{\rm E}^3=16
\sqrt{3}/27$, via Eqs.\,(\ref{eq:assi5})-(\ref{eq:upe5}).

\begin{figure*}[t]  
\begin{center}      
\includegraphics[scale=0.8]{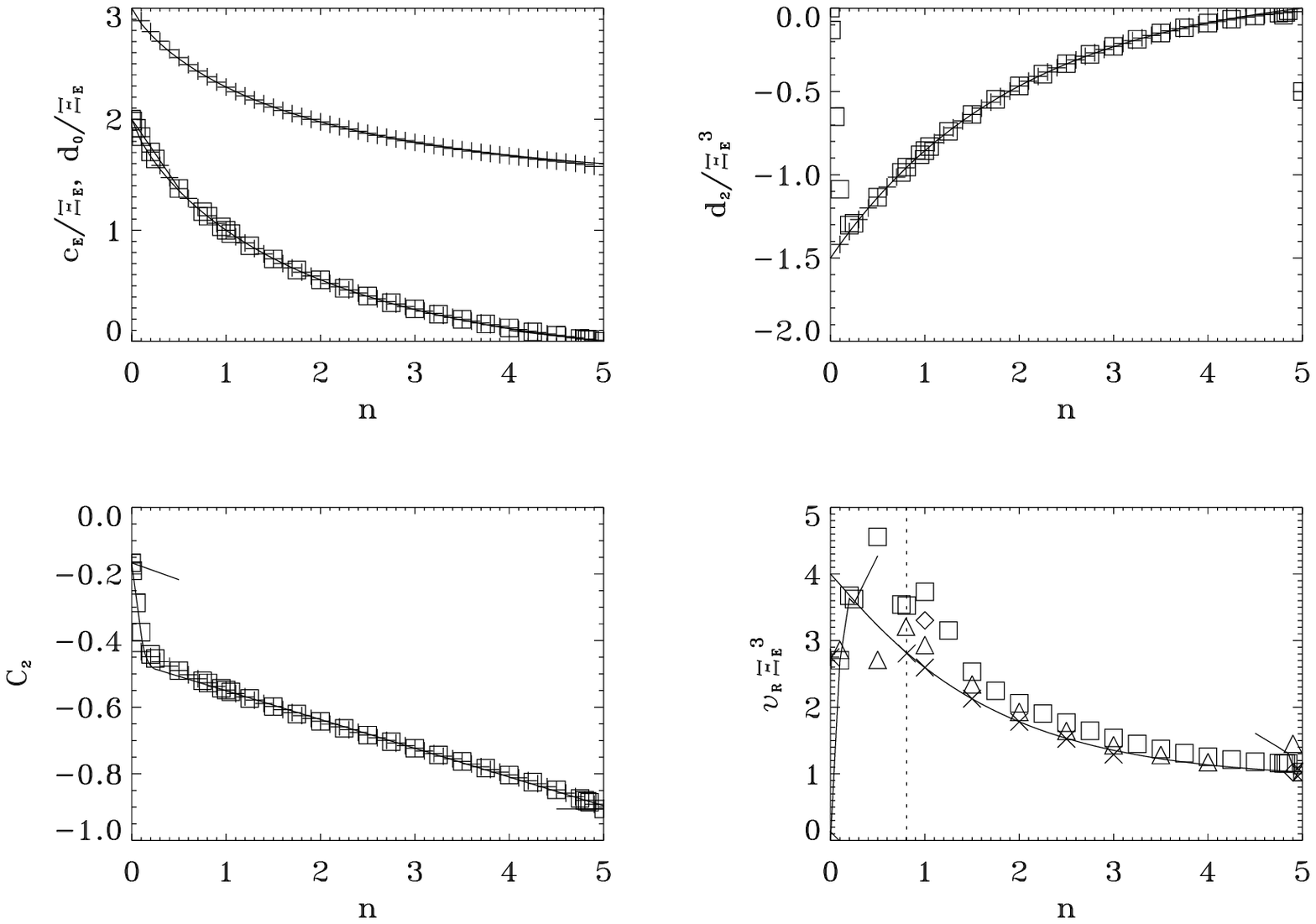}                      
\caption[ddbb]{Parameters of rigidly rotating polytropes,
$c_{\rm E}/\Xi_{\rm E}$ (squares), $d_0/\Xi_{\rm E}$ (crosses), top left,
$d_2/\Xi_{\rm E}^3$, top right, $C_2$, bottom left,
$\upsilon_{\rm R}\Xi_{\rm E}^3$, bottom right, as a function of the polytropic
index, $n$.  Source of results: squares (C85); crosses (H90); triangles
(Horedt 1983); saltires (James 1964); diamonds C80 or quoted therein or quoted
elsewhere (Horedt 2004); asterisks (Jeans 1929).   Curves represent simple
fits to the
data.   Series approximation are shown as broken lines.   The vertical
dotted line on the bottom right panel marks the boundary between instability 
towards bar modes (left) and equatorial breakup (right).   See text for
further details}.
\label{f:accv4}     
\end{center}       
\end{figure*}

The parameters under consideration are plotted as a function of the polytropic
index, $n$, in Fig.\,\ref{f:accv4}, top left panel ($c_{\rm E}/\Xi_{\rm E}$,
lower case; $d_0/\Xi_{\rm E}$, upper case), top right panel
($d_2/\Xi_{\rm E}^3$), bottom left panel ($C_2$), bottom right panel
($\upsilon_{\rm R}\Xi_{\rm E}^3$), where different symbols denote results
from different sources, in particular squares (C85) and crosses (H90).
Related interpolation curves are also shown together with analytical
approximations in the neighbourhood of $n=0,5$, within the range,
$0\le n\le0.5$, $5\ge n\ge4.5$, respectively.   For a formal derivation and
further details, an interested reader is addressed to Appendix \ref{a:defc}.
The transition between instability towards bar modes (left) and equatorial
breakup (right), $n=0.808$, (James 1964), is marked by a dotted vertical line
on the bottom right panel of Fig.\,\ref{f:accv4}.

With regard to an assigned empirical function, $f(n)$, and related fitting
curve, $f_{\rm fit}(n)$, the relative error may be expressed as:
\begin{lefteqnarray}
\label{eq:erre}
&& R[f(n)]=1-\frac{f_{\rm fit}(n)}{f(n)}~~;
\end{lefteqnarray}
which 
\begin{figure*}[t]  
\begin{center}      
\includegraphics[scale=0.8]{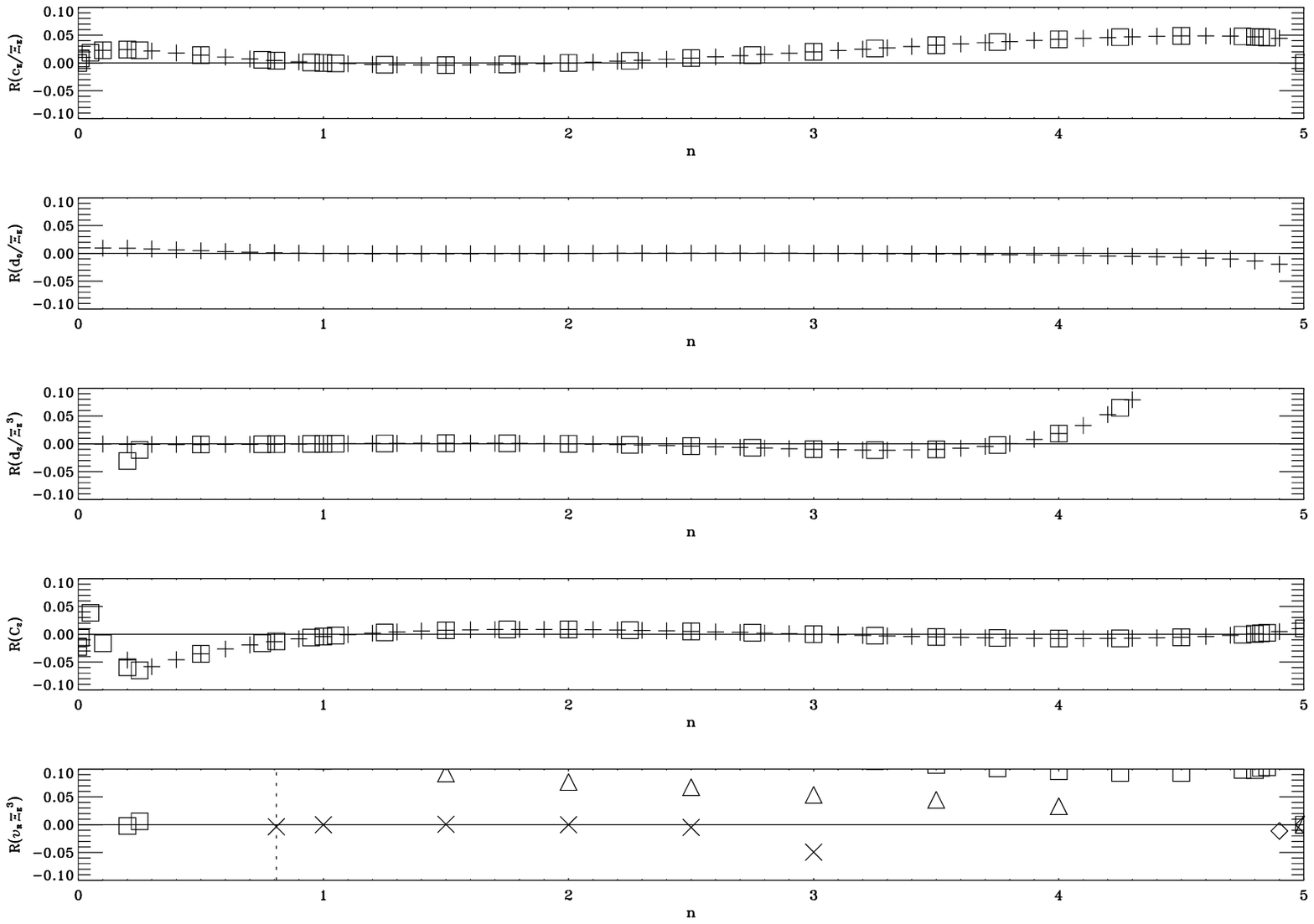}                      
\caption[ddbb]{The relative error, $R[f(n)]=1-f_{\rm fit}(n)/f(n)$, as a
function of the polytropic index, $n$, for (from top to bottom) $f(n)=
c_{\rm E}/\Xi_{\rm E}$, $d_0/\Xi_{\rm E}$,
$d_2/\Xi_{\rm E}^3$, $C_2$,
$\upsilon_{\rm R}\Xi_{\rm E}^3$, where $f_{\rm fit}$ relates to the fitting
curve and $f$ to results from different sources plotted with the same symbols
as in Fig.\,\ref{f:accv4}.}
\label{f:dccv4}     
\end{center}       
\end{figure*}                                                                     
is plotted in Fig.\,\ref{f:dccv4} for the cases of interest (from top to
bottom), $f(n)=c_{\rm E}/\Xi_{\rm E}$, $d_0/\Xi_{\rm E}$, $d_2/\Xi_{\rm E}^3$,
$C_2$, $\upsilon_{\rm R}\Xi_{\rm E}^3$, respectively, where different symbols
denote results from different sources, in particular squares (C85) and crosses
(H90).

An inspection of Figs.\,\ref{f:accv4} and \ref{f:dccv4} discloses the
following.
\begin{description}
\item[$\bullet$\hspace{6mm}]
The parameter, $c_{\rm E}/\Xi_{\rm E}$, matches to the limiting case, $n=0$,
for both C85 and H90 results.   The parameters, $d_0/\Xi_{\rm E}$,
$d_2/\Xi_{\rm E}^3$, $C_2$, $\upsilon_{\rm R}\Xi_{\rm E}^3$, match to the
limiting case, $n=0$, for C85 ($d_0/\Xi_{\rm E}$ unavailable) results, while
the contrary holds for H90 ($\upsilon_{\rm R}\Xi_{\rm E}^3$ unavailable)
results.   The discrete domain is $n=\Delta n\,i$, where $\Delta n=0.25$,
$0\le i\le20$, with additional points in the neighbourhood of $n=0,5,$ for C85
results and $\Delta n=0.10$, $0\le i\le50$, for H90 results.
\item[$\bullet$\hspace{6mm}]
The parameters,  $c_{\rm E}/\Xi_{\rm E}$, $C_2$,
$\upsilon_{\rm R}\Xi_{\rm E}^3$, match to the limiting case, $n=5$, for both
C85 and H90 ($\upsilon_{\rm R}\Xi_{\rm E}^3$ unavailable) results, while the
contrary holds with regard to the parameters, $d_0/\Xi_{\rm E}$,
$d_2/\Xi_{\rm E}^3$, for both C85 ($d_0/\Xi_{\rm E}$ unavailable) and H90
results.
\item[$\bullet$\hspace{6mm}]
The parameter, $\upsilon_{\rm R}\Xi_{\rm E}^3$, exhibits a nonmonotonic trend
in fissional regime, $0\le n\le0.808$, for C85 results, even if the
discrepancy with respect to numerical computations is unacceptably large.
\item[$\bullet$\hspace{6mm}]
For $n\appgeq0$, $n\appleq5$, the analytical approximation provides an
excellent fit to the results in connection with $c_{\rm E}/\Xi_{\rm E}$, no
acceptable fit but a right trend in connection with $C_2$,
$\upsilon_{\rm R}\Xi_{\rm E}^3$, no acceptable fit and a worse trend (not
shown) in connection with $d_0/\Xi_{\rm E}$, $d_2/\Xi_{\rm E}^3$.
\item[$\bullet$\hspace{6mm}]
The relative errors of the fitting curves lie within about 5\% for both C85
and H90 results in connection with the parameters, $c_{\rm E}/\Xi_{\rm E}$,
$d_0/\Xi_{\rm E}$, $d_2/\Xi_{\rm E}^3$ ($n\appleq4$), $C_2$, and for results
from numerical computations, (James 1964), in connection with the parameter,
$\upsilon_{\rm R}\Xi_{\rm E}^3$.   A larger discrepancy related to
$d_2/\Xi_{\rm E}^3$ ($n\appgeq4$), is due to increasingly smaller values
as $n\to5$ via Eq.\,(\ref{eq:erre}).
\end{description}

\section{Discussion}
\label{disc}

Open questions on the theory of EC polytropes have been focused as
\begin{description}
\item[(1)\hspace{2mm}]
What about the validity of EC associated equations, which imply the following
approximations:
\begin{lefteqnarray*}
&& \theta^n=[\theta_0+A_2\theta_2P_2(\mu)]^n\approx\theta_0^n+n\theta_0^{n-1}
A_2\theta_2P_2(\mu)~~; \\
&& \theta^n=\{\theta_{\rm E}+\upsilon[\psi_0+C_2\psi_2P_2(\mu)]\}^n
\approx
\theta_{\rm E}^n+\upsilon n\theta_{\rm E}^{n-1}[\psi_0+C_2\psi_2P_2(\mu)]~~;
\end{lefteqnarray*}
on a region sufficiently close to the boundary,
where $\theta_0\approx0$ and $\theta_{\rm E}\approx0$?
\item[(2)\hspace{2mm}]
What about the continuity of selected parameters as function of the polytropic
index, $n$, in the close neighbourhood of the limiting cases, $n=0,5$?
\end{description}
Concerning (1), $\theta(\xi_0,\mu_0)=\theta_0(\xi_0)$ via Eqs.\,(\ref{eq:iso})
and (\ref{eq:is0}), within the framework of the extended EC approximation.
Then the EC equation, Eq.\,(\ref{eq:ECa}), for $\mu=\mu_0$ reduces to the EC
associated equation of degree, $2\ell=0$, Eq.\,(\ref{eq:EC2l}), via
Eqs.\,(\ref{eq:Le}), (\ref{eq:R10}), and the related solution holds all over
the radially distorted sphere.   On the other hand, Eq.\,(\ref{eq:R10})
implies the validity of the EC associated equation of order, $2\ell=2$,
Eq.\,(\ref{eq:EC2l}), provided $\mu$ is close enough to $\mu_0$.   Then the
related solution holds all over the radially distorted sphere.
In fact, the EC associated functions, $\theta_{2\ell}$, depend only on the
radial coordinate and remain unchanged along an arbitrary spherical surface
centered on the origin.

This is why, though the validity of the EC
associated equations appears restricted to a special region within the system,
still their solutions hold at any point, provided $\theta_0^n$ is the real
part of the principal value of the complex power within the range,
$\Xi_0<\xi<\Xi_e$, (Linnel 1981; C83; Geroyannis 1988; Geroyannis and
Karageorgopoulos 2014).   More specifically, $\theta_{2\ell}$
must be conceived as the solutions of EC associated equations in the region
where Eq.\,(\ref{eq:thn2l}) holds to a good extent, and solutions of different
versions of EC associated equations in the region where Eq.\,(\ref{eq:thn2l})
is an unacceptable approximation.

The above results are valid for the EC associated functions, $\theta_0$ and
$\theta_2$, in the framework of the extended C33 approximation.   The same
holds for the EC associated functions, $\psi_0$, $\psi_2$, via
Eqs.\,(\ref{eq:psig0})-(\ref{eq:psig2}), in the framework of the C33
approximation.

Concerning (ii), among the selected parameters, the comparison between results
from different sources (C85; H90) can be performed only for
$c_{\rm E}/\Xi_{\rm E}$, $d_2/\Xi_{\rm E}^3$, $C_2$.   An inspection of
Fig.\,\ref{f:accv4} shows related results agree to a good extent for $n>0.1$,
while H90 results are lacking within the range, $0<n<0.1$.   The sole
significant discrepancy occurs for $d_2/\Xi_{\rm E}^3$ at $n=0.1$.  The reason
can be found in the parent papers, where the EC associated functions,
$\psi_0$, $\psi_2$, and their first derivatives, match to their counterparts
at $n=0$ in one case (C85; $\psi_0$ unavailable), while the contrary
holds in the other one (H90).   The continuity is implied in that $\psi_0$,
$\psi_2$, are related to the gravitational potential inside the body, which
is expected to change continuously as the density profile tends to be flat.

To this respect, it is worth emphasizing the above mentioned attempts use
different methods for solving the EC associated equations, namely series
solutions (C85) and direct numerical integration (H90).   The results are in
perfect agreement with regard to the EC function and its first derivatives.
The same holds for the EC associated function,
$\theta_2(\Xi_{\rm E};\upsilon=0)=\psi_2(\Xi_{\rm E})$ ($\psi_0$ has not been
evaluated in C83)
provided $n\ge1$.   On the contrary, significant discrepancies arise for 
$\theta_2(\Xi_{\rm E};\upsilon=0)=\psi_2(\Xi_{\rm E})$,
$\theta_2^\prime(\Xi_{\rm E};\upsilon=0)=\psi_2^\prime(\Xi_{\rm E})$, for
$n<0.2$, $n<0.5$, respectively, while
$\theta_2^\pprime(\Xi_{\rm E};\upsilon=0)=\psi_2^\pprime(\Xi_{\rm E})$ is
divergent
within the range, $0<n<0.5$, $0.5<n<1$, (C83).   It can be seen no exact value
of $\theta_{2\ell}(\Xi_{\rm E};\upsilon=0)=\psi_{2\ell}(\Xi_{\rm E})$ can be
obtained from numerical integration in the
neighbourhood of $n=0$ (H90).   Accordingly, results inferred via series
solutions must be preferred in this region, which ensures continuity (C83).
Then the selected parameters may safely be thought of as continuous as the
polytropic index, $n$, approaches 0 i.e. the system tends to be homogeneous.

In the opposite limit, $n\to5$, the system tends to be infinitely extended or
infinitely concentrated.   The parameters, $c_{\rm E}/\Xi_{\rm E}$, $C_2$,
$\upsilon_{\rm R}\Xi_{\rm E}^3$, may safely be thought of as continuous,
while the contrary holds for $d_0/\Xi_{\rm E}$, $d_2/\Xi_{\rm E}^3$, where the
trend remains monotonic as shown in Fig.\,\ref{f:accv4}.   To this respect,
two orders of considerations can be drawn.   First, computations in the
neighbourhood of $n=5$ should be very accurate due to the divergence of
$\Xi_{\rm E}$, which maximizes the effect of the errors.   Second,
computations should be extended within the range, $4.9<n<5$, to recognize if
the expected trend takes place.

\section{Conclusion}
\label{conc}

With regard to polytropes,
the isopycnic surfaces can be approximated as similar and similarly placed
ellipsoids (exact for homogeneous configurations) for several
investigations, such as the description of
gravitational radiation from collapsing and rotating massive star cores,  
(Saens and Shapiro 1978, 1981), rigidly rotating and binary polytropes,
(Lai et al. 1993, 1994a,b), gravitational collapse of nonbaryonic dark matter
and related pancake formation, (Bisnovatyi-K\"ogan 2004, 2005).

In addition, inhomogeneous configurations sufficiently close to the extended
Roche limit $(n\appleq5)$ can be described, to an acceptable
extent, in terms of properties related to extended Roche configurations
$(n=5)$.   A description of tenuous gas-dust atmospheres of some stars
and tenuous haloes surrounding compact elliptical galaxies, in terms of
extended Roche configurations, is mentioned in a recent investigation
(Kondratyev and Trubitsina 2013).
  
On the other hand, numerical simulations have not been performed
(to the knowledge of the author) outside the range, $0.1\le n\le4.9$.   Then
a first step in exploiting the limits, $n\to0$, $n\to5$, with regard to
selected physical parameters, must necessarily be performed analytically.

The current attempt has been devoted to two specific points about EC
polytropes, namely (i) the extent to which both C33 and extended C33
approximation are consistent with the binomial series approximation,
$(\theta_w+\Delta\theta)^n\approx\theta_w^n+n\theta_w^{n-1}
\Delta\theta$, keeping in mind $\theta_w\to0$ on the boundary, $w={\rm E},0$,
respectively, and (ii) the trend shown by selected parameters
as a function of the polytropic index, $n$, in particular the continuity in
the neighbourhood of the limiting cases, $n=0,5$.   The main results may be
summarized as follows.
\begin{description}
\item[$\bullet$\hspace{6mm}]
Though the validity of EC associated equations is restricted to a specific
region within the system, still related solutions hold within the whole volume
with regard to the EC associated functions, $\theta_0$, $\theta_2$, in the
framework of the extended C33 approximation, and $\psi_0$, $\psi_2$, in the
framework of the C33 approximation.
\item[$\bullet$\hspace{6mm}]
The expected continuity of the parameters, $c_{\rm E}/\Xi_{\rm E}$,
$d_0/\Xi_{\rm E}$, $d_2/\Xi_{\rm E}^3$, $C_2$,
$\upsilon_{\rm R}\Xi_{\rm E}^3$, as a function of the polytropic index, $n$,
has been safely verified as $n\to0$ with the exception of $d_0/\Xi_{\rm E}$,
$\upsilon_{\rm R}\Xi_{\rm E}^3$, and as $n\to5$ with the exception of
$d_0/\Xi_{\rm E}$, $d_2/\Xi_{\rm E}^3$, where additional data would be needed.
\item[$\bullet$\hspace{6mm}]
Simple fits to the above mentioned functions are provided for a wide range of
$n$, where the relative error does not exceed a few percent.   Related curves
are exponential depending on four parameters, with the exception of $C_2$,
where two straight lines are joined by a parabolic segment.
\end{description}
The expression of physical parameters in terms of the polytropic index, $n$,
can be used in building up sequences of configurations with changing density
profile for assigned mass and angular momentum.


\appendix
\section*{Appendix}

\section{Slowly rotating isopycnic surfaces}
\label{a:P2}

\begin{figure*}[t]  
\begin{center}      
\includegraphics[scale=0.8]{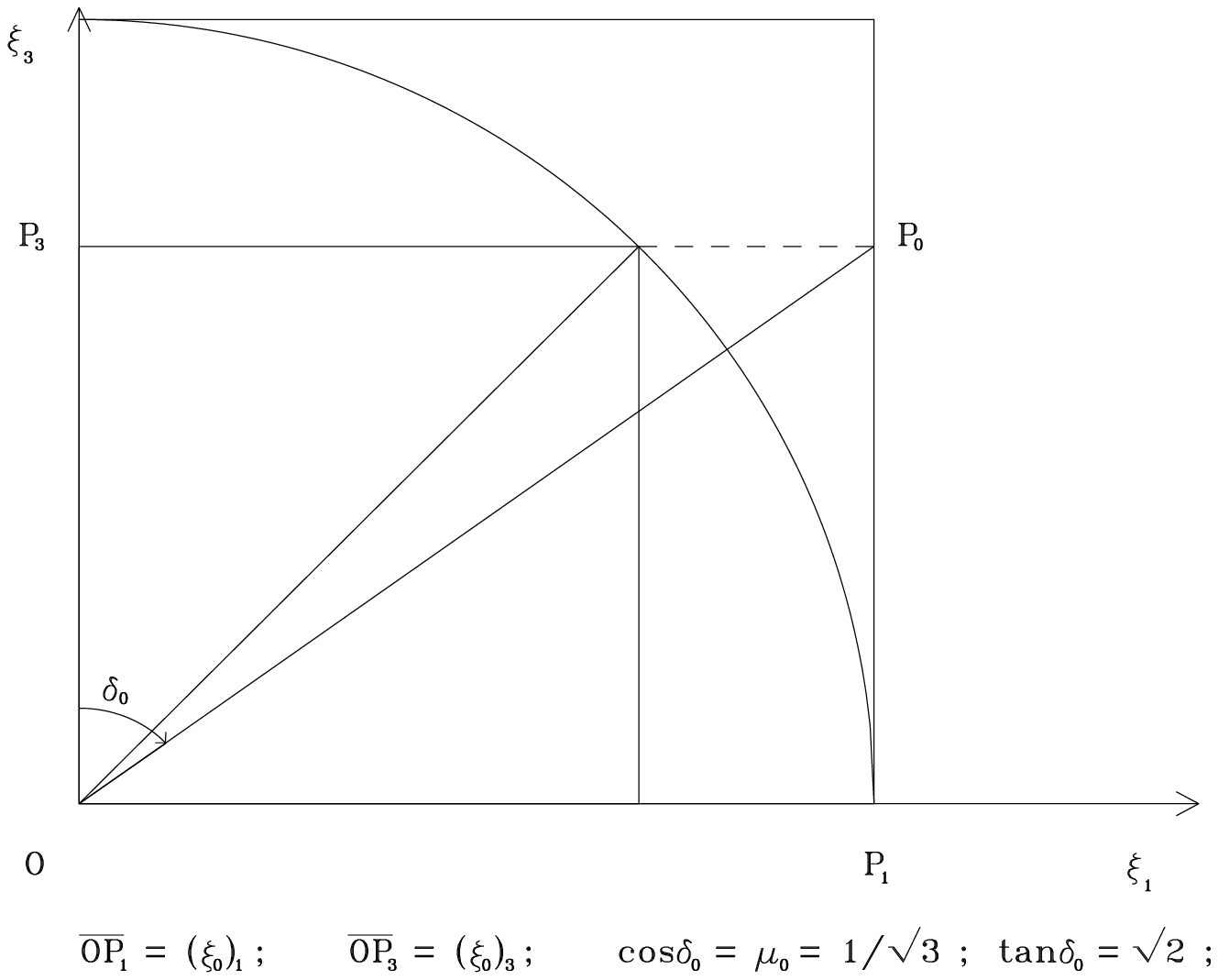}                      
\caption[ddbb]{Determination of the intersection point, ${\sf P}_0\equiv
(\xi_0,+\mu_0)\equiv[(\xi_0)_1,(\xi_0)_3]$, $(\xi_0)_1=\xi_0\sqrt{1-\mu_0^2}$,
$(\xi_0)_3=\xi_0\mu_0$, between fictitious (spherical)
isopycnic surfaces, $\theta_0(\xi_0)=\kappa$,
and rigidly rotating (oblate) isopycnic surfaces, $\theta(\xi,\mu)=\kappa$,
and hence the locus of intersection points, $(\xi_0,\mp\mu_0)$, in the limit
of small rotation.
(i) Fix $(\xi_0)_3$ and trace the square with side equal to
$(\xi_0)_3$ and three vertexes on the non negative coordinate semiaxes.
(ii) Trace the quarter of circle
centered on the origin, with radius equal to the diagonal of the square
defined in (i), lying on the first quadrant.
(iii) Trace the square with
side equal to the diagonal of the square defined in (i) and three vertexes on
the non negative coordinate semiaxes.
(iv) Trace the continuation of the horizontal side of the square defined in
(i), parallel to the horizontal axis, up to the intersection
with the vertical side of the square defined in (iii), parallel to the
vertical axis.
(v) The intersection defined in (iv) yields the point,
${\sf P}_0\equiv(\xi_0,+\mu_0)\equiv[(\xi_0)_1,(\xi_0)_3]$, and hence the
locus, $(\xi_0,\mp\mu_0)$.}
\label{f:iso2}     
\end{center}       
\end{figure*}

In the limit of small rotation, $\upsilon\ll1$, it may safely be thought the
series, expressed by Eq.\,(\ref{eq:iso}), reduces to:
\begin{lefteqnarray}
\label{eq:isl}
&& \theta(\xi,\mu)=\theta_0(\xi)+A_2\theta_2(\xi)P_2(\mu)=\kappa~~; \\
\label{eq:P2}
&& P_2(\mu)=\frac32\mu^2-\frac12~~;
\end{lefteqnarray}
which implies $\theta(\xi_0,\mp\mu_0)=\theta_0(\xi_0)$, where $\mu_0=\cos
\delta_0=1/\sqrt{3}$, $\delta_0=\arctan\sqrt{2}$.   The dimensionless radius,
$\xi_0=\sqrt{[(\xi_0)_1]^2+[(\xi_0)_3]^2}$,
$(\xi_0)_1=\xi_0\sqrt{1-\mu_0^2}$, $(\xi_0)_3=\mp\xi_0\mu_0$,
may be geometrically determined in an
elegant way, as shown in Fig.\,\ref{f:iso2}, along the following steps.
\begin{description}
\item[$\bullet$\hspace{6mm}]
Fix $(\xi_0)_3=\xi_0\mu_0$ and trace the square with side equal to
$(\xi_0)_3$ and three vertexes on the non negative coordinate semiaxes.
\item[$\bullet$\hspace{6mm}]
Trace the quarter of circle
centered on the origin, with radius equal to the diagonal of the square
defined above, lying on the first quadrant.
\item[$\bullet$\hspace{6mm}]
Trace the square with
side equal to the diagonal of the square defined above and three vertexes on
the non negative coordinate semiaxes.
\item[$\bullet$\hspace{6mm}]
Trace the continuation of the horizontal side of the former
square defined above, parallel to the horizontal axis, up to the intersection
with the vertical side of the latter square defined above, parallel to the
vertical axis.
\item[$\bullet$\hspace{6mm}]
The intersection defined above yields the point,
${\sf P}_0\equiv(\xi_0,+\mu_0)\equiv[(\xi_0)_1,(\xi_0)_3]$, and hence the
locus, $(\xi_0,\mp\mu_0)$.
\end{description}

\section{Determination of fitting curves}
\label{a:defc}

With regard to the parameters of EC polytropes, plotted in
Fig.\,\ref{f:accv4}, fitting curves shall be determined aiming to get simple
expressions instead of best fits.   To this end, a lot of symbols is needed,
part of which has already been used throughout the text with a different
meaning.   The reader has to keep in mind that symbols denoting parameters of
the fitting curves have no connection with their counterparts (if any) defined
in the text.

In the neighbourhood of polytropic
indexes where exact analytic results can be determined, $n=0,1,5$, a series
approximation can be developed (Seidov and Kuzackhmedov 1978; C88).
Unfortunately, for $n=0,5$, the quantities of interest cannot be
satisfactorily fitted, as shown by the broken lines (if any) plotted in
Fig.\,\ref{f:accv4}.   The sole exception is $c_{\rm E}/\Xi_{\rm E}$, where an
excellent fit is provided within the range, $0<n<0.5$, $4<n<5$, as shown in
Fig.\,\ref{f:accv4}, top left panel, lower case.

For $n\appgeq0$, the result is:
\begin{lefteqnarray}
\label{eq:prm10}
&& \frac{c_{\rm E}}{\Xi_{\rm E}}=2[1+(6\ln2-5)n]~~; \\
\label{eq:prm20}
&& \frac{d_0}{\Xi_{\rm E}}=\left(8\ln2-\frac{37}3\right)n~~; \\
\label{eq:prm30}
&& \frac{d_2}{\Xi_{\rm E}^3}=0~~; \\
\label{eq:prm40}
&& C_2=-\frac16\left[1+\frac{18}{25}\left(\frac{23}{15}-\ln2\right)n\right]~~;
\\
\label{eq:prm50}
&& \upsilon_{\rm R}\Xi_{\rm E}^3=\upsilon_{\rm R}(n)6\sqrt6\left[1+\left
(6\ln2-\frac72\right)n\right]~~; \\
\label{eq:prm60}
&& \Xi_{\rm E}=\sqrt6\left[1+\left(2\ln2-\frac76\right)n\right]~~;
\end{lefteqnarray}
where $\upsilon_{\rm R}$ is the value of the rotation parameter at the onset
of equatorial breakup (leaving aside instabilities) for EC polytropes of
index, $n$.   In the limit,
$n\to0$, Eqs.\,(\ref{eq:prm10})-(\ref{eq:prm40}) reduce to their exact
counterparts, Eq.\,(\ref{eq:par0}).

For $n\appleq5$, the result is: 
\begin{lefteqnarray}
\label{eq:prm15}
&& \frac{c_{\rm E}}{\Xi_{\rm E}}=\frac\pi{32}(5-n)~~; \\
\label{eq:prm25}
&& \frac{d_0}{\Xi_{\rm E}}=0~~; \\
\label{eq:prm35}
&& \frac{d_2}{\Xi_{\rm E}^3}=-\frac12~~; \\
\label{eq:prm45}
&& C_2=-\frac{128}{45\pi}~~;  \\
\label{eq:prm55}
&& \upsilon_{\rm R}\Xi_{\rm E}^3=\upsilon_{\rm R}(n)\left(\frac{32\sqrt3}\pi
\right)^3\frac1{(5-n)^3}~~; \\
\label{eq:prm65}
&& \Xi_{\rm E}=\sqrt3\left\{\left[\frac{32}{\pi(5-n)}\right]^2-1\right\}^{1/2}
\approx\frac{32\sqrt3}{\pi(5-n)}~~;
\end{lefteqnarray}
which coincide with or reduce to their exact counterparts,
Eq.\,(\ref{eq:par5}), as $n\to5$.

For $n\approx1$, a regular trend is shown by the data plotted in
Fig.\,\ref{f:accv4} and no analytical approximation is needed to gain further
insight.

With regard to the parameter, $C_2$, the fit has been performed in the
following way.   First, regression lines have been determined,
using standard methods, close enough to and far enough from $n=0$,
respectively.   The result is:
\begin{lefteqnarray}
\label{eq:intU}
&& C_2=a_{\rm U}n+b_{\rm U}~~;\qquad{\rm U=F,C}~~; \\
\label{eq:abF}
&& a_{\rm F}=-2.10475~~;\qquad b_{\rm F}=-0.170428~~;\qquad0\le n\le0.1~~; \\
\label{eq:abC}
&& a_{\rm C}=-0.0863467~~;\qquad b_{\rm C}=-0.464166~~;\qquad0.808\le n\le5~~;
\end{lefteqnarray}
as shown in Fig.\,\ref{f:accv4}, bottom left panel.   The indexes, F, C,
denote instability with respect to bar modes ($n<0.808$) and equatorial
breakup ($n\ge0.808$), respectively.   From this point on, in the case under
discussion, the notation, $({\sf O}nC_2)=({\sf O}xy)$, shall be used for
simplicity.

Second, the straight lines are joined by a parabola requiring the following.
\begin{description}
\item[$\bullet$\hspace{6mm}]
The axis of the parabola coincides with the bisector of the angle formed by
the regression lines, expressed as:
\begin{lefteqnarray}
\label{eq:amga}
&& \gamma=\arctan\frac{a_{\rm C}-a_{\rm F}}{1+a_{\rm C}a_{\rm F}}~~;
\end{lefteqnarray}
according to standard results of analytic geometry.
\item[$\bullet$\hspace{6mm}]
The first derivatives of the joining parabola, $p(x,y)=0$, at the joining
points, ${\sf P}_{\rm F}\equiv(x_{\rm F}, y_{\rm F})$,
${\sf P}_{\rm C}\equiv(x_{\rm C}, y_{\rm C})$, equal the slope of related
regression lines, as:
\begin{lefteqnarray}
\label{eq:dpU}
&& \left(\frac{\partial p}{\partial x}\right)_{{\sf P}_{\rm U}}=
a_{\rm U}~~;\qquad{\rm U=F,C}~~;
\end{lefteqnarray}
or, in other words, the joining parabola is tangent to the regression
lines at the points, ${\sf P}_{\rm F}$, ${\sf P}_{\rm C}$, respectively.
\end{description}

For determining the equation of the parabola, a change of reference frame is
needed, where the new origin, ${\sf Q}$, coincides with the intersection point
between the regression lines, ${\sf P}_{\rm I}\equiv(x_{\rm I}, y_{\rm I})$,
the vertical
axis, $Y$, coincides with the bisector of the angle, $\gamma$, formed by
the regression lines, and the chirality is preserved i.e. $({\sf O}xy)$
and $({\sf Q}XY)$ can be superimposed on the common plane.

The explicit expression of the coordinates of the intersection point reads:
\begin{lefteqnarray}
\label{eq:Pi}
&& x_{\rm I}=-\frac{b_{\rm F}-b_{\rm C}}{a_{\rm F}-a_{\rm C}}~~;\qquad y_{\rm I}=\frac
{a_{\rm F}b_{\rm C}-a_{\rm C}b_{\rm F}}{a_{\rm F}-a_{\rm C}}~~;
\end{lefteqnarray}
according to standard results of analytic geometry.

Finally, the change of reference frame takes the expression:
\begin{lefteqnarray}
\label{eq:QXY}
&&
\cases{
X=(x-x_{\rm I})\cos\theta-(y-y_{\rm I})\sin\theta~~; \cr
Y=(x-x_{\rm I})\sin\theta+(y-y_{\rm I})\cos\theta~~; \cr
}
\end{lefteqnarray}
where $\theta$ is the angle between the starting axis, $y$, and the resulting
axis, $Y$, i.e. the bisector of the angle, $\gamma$.

The related explicit expression reads:
\begin{lefteqnarray}
\label{eq:teta}
&& \theta=-\frac12(\arctan a_{\rm F}+\arctan a_{\rm C})~~; 
\end{lefteqnarray}
according to standard results of analytic geometry.

With regard to the resulting reference frame, $({\sf Q}XY)$, the regression
lines intersect at the origin, ${\sf Q}\equiv{\sf P}_{\rm I}$, and exhibit
equal and opposite slopes as:
\begin{lefteqnarray}
\label{eq:ap}
&& a^\prime=-a_{\rm F}^\prime=a_{\rm C}^\prime=-\tan\frac\gamma2=\tan\left(
\frac12\arctan\frac{a_{\rm C}-a_{\rm F}}{1+a_{\rm C}a_{\rm F}}\right)~;
\end{lefteqnarray}
according to standard results of analytic geometry.

In addition, the joining points, ${\sf P}_{\rm F}\equiv(X_{\rm F},Y_{\rm F})$,
${\sf P}_{\rm C}\equiv(X_{\rm C},Y_{\rm C})$, are symmetrical with respect to
the $Y$ axis as:
\begin{lefteqnarray}
\label{eq:XYsi}
&& -X_{\rm F}=X_{\rm C}~~;\qquad Y_{\rm F}=Y_{\rm C}~~;
\end{lefteqnarray}
and the equation of the parabola reduces to:
\begin{lefteqnarray}
\label{eq:par}
&& Y=P(X)=AX^2+C~~;
\end{lefteqnarray}
where the term in
$X$ is ruled out by the above mentioned symmetry.

For assigned $(x_{\rm F},y_{\rm F})$ {\it or} $(x_{\rm C},y_{\rm C})$,
$(X_{\rm F},Y_{\rm F})$ {\it and} $(X_{\rm C},Y_{\rm C})$ can be determined
via Eqs.\,(\ref{eq:Pi})-(\ref{eq:XYsi}).   The remaining point,
$(x_{\rm C},y_{\rm C})$ or $(x_{\rm F},y_{\rm F})$, can be determined
inverting Eq.\,(\ref{eq:QXY}) as:
\begin{lefteqnarray}
\label{eq:Oxy}
&&
\cases{
x=x_{\rm I}+X\cos\theta+Y\sin\theta~~; \cr
y=y_{\rm I}-X\sin\theta+Y\cos\theta~~; \cr
}
\end{lefteqnarray}
and particularizing to $(x,y)=(x_{\rm C},y_{\rm C})$ or
$(x,y)=(x_{\rm F},y_{\rm F})$; $(X,Y)=(X_{\rm C},Y_{\rm C})$ or
$(X,Y)=(X_{\rm F},Y_{\rm F})$; respectively.

The coefficients, $A$, $C$, appearing in Eq.\,(\ref{eq:par}), can be expressed
keeping in mind the joining parabola is tangent to the regression lines
at $(X_{\rm F},Y_{\rm F})$ and $(X_{\rm C},Y_{\rm C})$.   The result is:
\begin{lefteqnarray}
\label{eq:Apa}
&& A=-\frac12\frac{a^\prime}{X_{\rm F}}=\frac12\frac{a^\prime}{X_{\rm C}}~~;
\\
\label{eq:Cpa}
&& C=Y_{\rm F}+\frac12\frac{a^\prime}{X_{\rm F}}=Y_{\rm C}-\frac12\frac
{a^\prime}{X_{\rm C}}~~;
\end{lefteqnarray}
and the equation of the parabola in the starting reference frame, $p(x,y)=0$,
can be determined inserting Eqs.\,(\ref{eq:QXY}), (\ref{eq:Apa}),
(\ref{eq:Cpa}), into (\ref{eq:par}).

In summary, the above mentioned procedure acts along the following steps.
\begin{description}
\item[(a)\hspace{2mm}]
Choose a joining point, ${\sf P}_{\rm U}\equiv(x_{\rm U},y_{\rm U})$, on the
regression line, $y=a_{\rm U}x+b_{\rm U}$, U = F or U = C.
\item[(b)\hspace{2mm}]
Change the reference frame from $({\sf O}xy)$ to $({\sf Q}XY)$.
\item[(c)\hspace{2mm}]
With regard to the resulting frame, $({\sf Q}XY)$, determine the joining
points, ${\sf P}_{\rm F}\equiv(X_{\rm F},Y_{\rm F})$,
${\sf P}_{\rm C}\equiv(X_{\rm C},Y_{\rm C})$, where $-X_{\rm F}=X_{\rm C}$,
$Y_{\rm F}=Y_{\rm C}$.
\item[(d)\hspace{2mm}]
With regard to the starting frame, $({\sf O}xy)$, determine the joining point,
${\sf P}_{\rm V}\equiv(x_{\rm V},y_{\rm V})$, V = C or V = F.
\item[(e)\hspace{2mm}]
With regard to the resulting frame, $({\sf Q}XY)$, determine the coefficients
in the expression of the parabola, Eq.\,(\ref{eq:par}).
\item[(d)\hspace{2mm}]
With regard to the starting frame, $({\sf O}xy)$, write the explicit
expression of the equation of the parabola, $p(x,y)=0$.
\end{description}

The parameters of the joining parabola are found to be:
\begin{lefteqnarray}
\label{eq:parpa1}
&& (x_{\rm F},y_{\rm F})=(0.1,0.380903)~~;\quad(x_{\rm C},y_{\rm C})=
(0.251231,0.485859)~~; \\ 
\label{eq:parpa2}
&& A=3.11439~~;\quad C=0.026384~~;
\end{lefteqnarray}
and the related fitting curve can be determined by use of
Eqs.\,(\ref{eq:intU}), (\ref{eq:abF}), (\ref{eq:abC}), (\ref{eq:parpa1}),
(\ref{eq:parpa2}), and
plotted in Fig.\,\ref{f:accv4}, bottom left panel.   The relative error,
$R(C_2)$, is shown in Fig.\,\ref{f:dccv4}, intermediate bottom panel.

With regard to the remaining parameters, $c_{\rm E}/\Xi_{\rm E}$,
$d_0/\Xi_{\rm E}$, $d_2/\Xi_{\rm E}^3$, $\upsilon_{\rm R}\Xi_{\rm E}^3$, a
fitting exponential function has been chosen, as:
\begin{lefteqnarray}
\label{eq:expf}
&& y=f(x)=C_1\exp(-C_2x^\gamma)+C_3~~;
\end{lefteqnarray}
where $x=n$, $y=f(x)$ is the parameter of interest, and $C_1$, $C_2$, $C_3$,
$\gamma$, are constants to be determined.   The related boundary conditions
are:
\begin{lefteqnarray}
\label{eq:ebc0}
&& f(0)=C_1+C_3=y_0~~; \\
\label{eq:ebc1}
&& f(1)=C_1\exp(-C_2)+C_3=y_1~~; \\
\label{eq:ebc2}
&& f(2)=C_1\exp(-C_22^\gamma)+C_3=y_2~~; \\
\label{eq:ebc5}
&& f(5)=C_1\exp(-C_25^\gamma)+C_3=y_5~~;
\end{lefteqnarray}
where $y_0$, $y_1$, $y_5$, can be analytically expressed and $y_2$ has to be
numerically computed.

The substitution of Eq.\,(\ref{eq:ebc0}) into (\ref{eq:expf}) after some
algebra yields:
\begin{lefteqnarray}
\label{eq:exC2}
&& \exp(-C_2)=\left(\frac{C_1-y_0+y}{C_1}\right)^{1/x^\gamma}~~;\qquad
\frac{C_1-y_0+y}{C_1}>0~~;
\end{lefteqnarray}
where the inequality must necessarily hold owing to (i) the exponential
function is always positive and (ii) the real basis of a power with a real
exponent is defined for non negative values.

Let $(x_{\rm A},y_{\rm A})$, $(x_{\rm B},y_{\rm B})$, $(x_{\rm C},y_{\rm C})$,
be generic points for
which the coordinates are known.   The particularization of
Eq.\,(\ref{eq:exC2}) to $(x_{\rm A},y_{\rm A})$, $(x_{\rm B},y_{\rm B})$,  and
the combination of related expressions, after some algebra yields:
\begin{lefteqnarray}
\label{eq:enBA}
&& \gamma\ln\frac{x_{\rm B}}{x_{\rm A}}=\ln\ln\frac{C_1-y_0+y_{\rm B}}{C_1}-
\ln\ln\frac{C_1-y_0+y_{\rm A}}{C_1}~~;
\end{lefteqnarray}
and the combination of Eq.\,(\ref{eq:enBA}) with its counterpart related to
$(x_{\rm A},y_{\rm A})$, $(x_{\rm C},y_{\rm C})$, after some algebra produces:
\begin{lefteqnarray}
\label{eq:eCBA}
&& \ln\frac{x_{\rm C}}{x_{\rm A}}\left[\ln\ln\frac{C_1-y_0+y_{\rm B}}{C_1}-
\ln\ln\frac{C_1-y_0+y_{\rm A}}{C_1}\right] \nonumber \\
&& =\ln\frac{x_{\rm B}}{x_{\rm A}}\left[\ln\ln\frac{C_1-y_0+y_{\rm C}}{C_1}-
\ln\ln\frac{C_1-y_0+y_{\rm A}}{C_1}\right]~~;
\end{lefteqnarray}
where the intersection of related curves yields the value of $C_1$.

In the case under discussion, $(x_{\rm A},y_{\rm A})=(1,y_1)$,
$(x_{\rm B},y_{\rm B})=(2,y_2)$, $(x_{\rm C},y_{\rm C})=(5,y_5)$, and
Eq.\,(\ref{eq:eCBA}) reduces to:
\begin{lefteqnarray}
\label{eq:e521}
&& \ln5\left[\ln\ln\frac{C_1-y_0+y_2}{C_1}-
\ln\ln\frac{C_1-y_0+y_1}{C_1}\right] \nonumber \\
&& =\ln2\left[\ln\ln\frac{C_1-y_0+y_5}{C_1}-
\ln\ln\frac{C_1-y_0+y_1}{C_1}\right]~~;
\end{lefteqnarray}
where the knowledge of $C_1$ together with $y_0$, $y_1$, $y_2 $, $y_5$, makes
the remaining parameters, $C_2$, $C_3$, $\gamma$, be inferred from
Eqs.\,(\ref{eq:ebc0})-(\ref{eq:ebc5}).   Related fitting curves are shown in
Fig.\,\ref{f:accv4}, top left, top right and bottom right panels.   For
further details, each case must be discussed separately.

With regard to the function, $f(n)=c_{\rm E}/\Xi_{\rm E}$, plotted in
Fig.\,\ref{f:accv4}, top left panel, lower case, the parameters of the fitting
curve are found to be:
\begin{lefteqnarray}
\label{eq:partl}
&& C_1=2.28169~;~~C_2=0.576739~;~~C_3=-0.281685~;~~\gamma=0.800546~;\qquad
\end{lefteqnarray}
related to the input parameters:
\begin{lefteqnarray}
\label{eq:inptl}
&& y_0=2~~;\quad y_1=1~~;\quad y_2=0.553897~~;\quad y_5=0~~;
\end{lefteqnarray}
and the relative error, $R[f(n)]$, is shown in Fig.\,\ref{f:dccv4}, extreme
top panel.

With regard to the function, $f(n)=d_0/\Xi_{\rm E}$, plotted in
Fig.\,\ref{f:accv4}, top left panel, upper case, the parameters of the fitting
curve are found to be:
\begin{lefteqnarray}
\label{eq:partd}
&& C_1=1.57191~;~~C_2=0.601048~;~~C_3=1.42809~;~~\gamma=0.809883~;\qquad
\end{lefteqnarray}
related to the input parameters:
\begin{lefteqnarray}
\label{eq:inptd}
&& y_0=3~~;\quad y_1=\frac{\pi^2}3-1\approx2.29987~~;\quad y_2=1.97614~~;\quad
y_5=1.6~~;\qquad
\end{lefteqnarray}
and the relative error, $R[f(n)]$, is shown in Fig.\,\ref{f:dccv4},
intermediate top panel.

With regard to the function, $f(n)=d_2/\Xi_{\rm E}^3$, plotted in
Fig.\,\ref{f:accv4}, top right panel, the parameters of the fitting curve are
found to be:
\begin{lefteqnarray}
\label{eq:partr}
&& C_1=-1.63794~;~~C_2=0.500459~;~~C_3=0.137937~;~~\gamma=0.993033~;\qquad
\end{lefteqnarray}
related to the input parameters:
\begin{lefteqnarray}
\label{eq:inptr}
&& y_0=-1.5~;\,~y_1=-\frac{15-\pi^2}6\approx-0.855066~;\,~y_2=-0.466983~;\,~
y_5=0~;\qquad
\end{lefteqnarray}
and the relative error, $R[f(n)]$, is shown in Fig.\,\ref{f:dccv4}, middle
panel.

With regard to the function, $f(n)=\upsilon_{\rm R}\Xi_{\rm E}^3$, plotted in
Fig.\,\ref{f:accv4}, bottom right panel, the parameters of the fitting curve
are found to be:
\begin{lefteqnarray}
\label{eq:parbr}
&& C_1=3.07748~;~~C_2=0.609298~;~~C_3=0.922519~;~~\gamma=1.06613~;\qquad
\end{lefteqnarray}
related to the input parameters:
\begin{lefteqnarray}
\label{eq:inpbr}
&& y_0=4~;~~y_1=2.59585~;~~y_2=1.78182~;~~y_5=\frac{16\sqrt{3}}{27}\approx
1.02640~;\qquad
\end{lefteqnarray}
and the relative error, $R[f(n)]$, is shown in Fig.\,\ref{f:dccv4}, extreme
bottom panel.

In addition, the combination of Eqs.\,(\ref{eq:prm65}), (\ref{eq:expf}),
(\ref{eq:parbr}), yields:
\begin{lefteqnarray}
\label{eq:upa5}
&& \upsilon_{\rm R}=\left[\frac{\pi(5-n)}{32\sqrt3}\right]^3[C_1\exp(-C_2n^
\gamma)+C_3]~~;
\end{lefteqnarray}
which is expected to hold to a good extent within the range, $4\le n\le5$,
avoiding computational errors due to large values of the dimensionless radius,
$\Xi_{\rm E}$.

\end{document}